\documentclass[12pt,aps,nofootinbib,prb,preprint,superscriptaddress,showpacs,showkeys,12pt,sort&compress]{revtex4}

\usepackage[T1]{fontenc}
\usepackage[latin9]{inputenc}
\usepackage{amsmath}
\usepackage{amssymb}
\usepackage{graphicx}
\usepackage{esint}
\PassOptionsToPackage{normalem}{ulem}
\usepackage{ulem}

\makeatletter
\@ifundefined{textcolor}{}
{%
 \definecolor{BLACK}{gray}{0}
 \definecolor{WHITE}{gray}{1}
 \definecolor{RED}{rgb}{1,0,0}
 \definecolor{GREEN}{rgb}{0,1,0}
 \definecolor{BLUE}{rgb}{0,0,1}
 \definecolor{CYAN}{cmyk}{1,0,0,0}
 \definecolor{MAGENTA}{cmyk}{0,1,0,0}
 \definecolor{YELLOW}{cmyk}{0,0,1,0}
}

\@ifundefined{date}{}{\date{}}

\usepackage{indentfirst}
\usepackage{amsfonts}
\usepackage[T1]{fontenc}
\usepackage{ae,aecompl}
\usepackage{sidecap}
\usepackage[section]{placeins}
\usepackage{epsfig}
\usepackage{epstopdf}
\date{\today}

\makeatother

\begin{document}

\title{Irreversible thermodynamics of creep in crystalline solids}

\author{\noindent Y. Mishin}

\address{School of Physics, Astronomy and Computational Sciences, MSN 3F3,
George Mason University, Fairfax, Virginia 22030, USA}

\author{\noindent J. A. Warren}

\address{\noindent Materials Science and Engineering Division, National Institute
of Standards and Technology, Gaithersburg, Maryland 20899, USA}

\author{\noindent R. F. Sekerka }

\address{Department of Physics, Carnegie Mellon University, Pittsburgh, Pennsylvania
15213, USA}

\author{\noindent W. J. Boettinger}

\address{\noindent Materials Science and Engineering Division, National Institute
of Standards and Technology, Gaithersburg, Maryland 20899, USA}
\begin{abstract}
We develop an irreversible thermodynamics framework for the description
of creep deformation in crystalline solids by mechanisms that involve
vacancy diffusion and lattice site generation and annihilation. The
material undergoing the creep deformation is treated as a non-hydrostatically
stressed multi-component solid medium with non-conserved lattice sites
and inhomogeneities handled by employing gradient thermodynamics.
Phase fields describe microstructure evolution which gives rise to
redistribution of vacancy sinks and sources in the material during
the creep process. We derive a general expression for the entropy
production rate and use it to identify of the relevant fluxes and
driving forces and to formulate phenomenological relations among them
taking into account symmetry properties of the material. As a simple
application, we analyze a one-dimensional model of a bicrystal in
which the grain boundary acts as a sink and source of vacancies. The
kinetic equations of the model describe a creep deformation process
accompanied by grain boundary migration and relative rigid translations
of the grains. They also demonstrate the effect of grain boundary
migration induced by a vacancy concentration gradient across the boundary.
\end{abstract}

\keywords{Irreversible thermodynamics, creep deformation, diffusion, lattice
sites, phase field}

\pacs{61.72.-y, 62.20.Hg, 65.40.-b, 66.30.-h}

\maketitle

\section{Introduction}

When subject to a high homologous temperature and a sustained mechanical
load below the yield strength, many materials exhibit a slow time-dependent
plastic deformation called creep. This mode of deformation has been
observed in different classes of materials ranging from metals and
alloys to ceramics, polymers and ice. While several creep deformation
mechanisms have been proposed over the years, we will focus in this
work on mechanisms that require creation and annihilation of lattice
sites.\cite{Stephenson1988} Such mechanisms include so-called diffusional
creep in which the deformation is mediated by vacancy diffusion through
the lattice (Nabarro-Herring creep)\cite{Nabarro1948,Herring1950}
or along grain boundaries (GBs) (Coble creep),\cite{Coble,Farghalli01}
as well as creep by dislocation climb. A number of other mechanisms
that do not necessarily involve site creation and annihilation, such
as the thermally activated dislocation glide, will not be considered
here.

Most of the models of creep developed so far have an \emph{ad hoc}
character: they are obtained by postulating a particular mechanism
and assuming a constitutive relation between the creep deformation
rate and a chosen driving force. The development of a general and
rigorous theory of creep deformation requires at least the following
three components: (i) a thermodynamic model of a mechanically stressed
crystalline solid with non-conserved lattice sites, (ii) a model of
microstructure evolution that includes redistribution of vacancy sinks
and sources and the motion of interfaces separating different phases
and/or grains, and (iii) a set of kinetic equations derived from the
entropy production rate\cite{De-Groot1984} and identification of
the appropriate set of fluxes (including the creep deformation rate)
and the conjugate driving forces. To our knowledge, a theory comprising
all three components has not been developed to date.

Several theories involving one or two of the above components can
be found in the literature. Svoboda \emph{et al.}\cite{Svoboda2006,Fischer2011}
proposed a creep model for multi-component alloys with a continuous
distribution of vacancy sinks and sources. By contrast to previous
work, their kinetic equations have not been simply postulated but
rather derived from the maximum dissipation principle. The authors
identified and clearly separated two components of the creep deformation
tensor, the volume dilation/contraction and the shear, and correctly
established their decoupled character. However, their thermodynamic
treatment of solid solutions is based on certain assumptions and approximations
that are not always justified. For example, they use the Gibbs-Duhem
equation which is valid only for hydrostatically stressed systems
and introduce so-called ``generalized'' chemical potentials which
include only the hydrostatic part of the stress tensor $\boldsymbol{\sigma}$
{[}see, e.g., Sekerka and Cahn\cite{Sekerka04} for criticism of using
only the hydrostatic part of $\boldsymbol{\sigma}$ (``solid pressure'')
in solid-state thermodynamics{]}. In view of the non-uniqueness of
chemical potentials of substitutional components in non-hydrostatic
solids\cite{Larche73,Larche_Cahn_78,Larche1985,Mullins1985,Sekerka04,Voorhees2004,Frolov2010d}
and the fact that the Gibbs free energy is no longer a useful thermodynamic
potential, development of thermodynamic models of stressed solids
should start from the first and second laws in the energy-entropy
representation\cite{Willard_Gibbs} and proceed with extreme care.

As such, a very general and rigorous thermodynamic treatments of multicomponent
solids was developed by Larché and Cahn\cite{Larche73,Larche_Cahn_78,Larche1985}
as an extension of Gibbs' thermodynamics\cite{Willard_Gibbs} to non-hydrostatic
solid systems. Although their analysis is valid for stressed solids
with any number of substitutional and interstitial components, it
relies of the assumption that the lattice sites are conserved. The
lattice conservation imposes the so-called ``network constraint''
which penetrates through all thermodynamic equations. It is assumed
that lattice sites can be created or destroyed only at defects such
as surfaces, interfaces and dislocation cores. Such defective regions
are excluded from the direct thermodynamic treatment and only enter
the theory through boundary conditions. Thus, the question of how
the vacancy sinks and sources operate is essentially left beyond the
theory. Mullins and Sekerka\cite{Mullins1985} proposed a similar
theory for multicomponent crystalline solids with a more general treatment
of point defects based on the concepts of extended variable sets.
Their theory assumes the conservation of lattice unit cells, which
is similar to the ``network constraint''. Both Larché and Cahn\cite{Larche73,Larche_Cahn_78,Larche1985}
and Mullins and Sekerka\cite{Mullins1985} analyzed equilibrium states
of the solid and did not study the irreversible thermodynamics of
creep deformation.%
\footnote{In Sect.~8.5, Larché and Cahn\cite{Larche1985} do discuss some creep
problems, but they treat creep through boundary conditions with perfect
site conservation inside the lattice.%
}

Furthermore, these thermodynamic theories of solids\cite{Larche73,Larche_Cahn_78,Larche1985,Mullins1985}
are purely ``classical'', in which all thermodynamic properties depend
only on local thermodynamic densities\cite{Willard_Gibbs} but not
their gradients. Accordingly, transition regions between different
phases are treated as geometric surfaces of discontinuity\cite{Willard_Gibbs}
endowed with certain postulated properties, such as the ability (or
inability) to support shear stresses or the capacity (or lack thereof)
to generate or absorb vacancies. Existing creep models\cite{Svoboda2006,Fischer2011}
are also classical and thus incapable of describing the microstructure
evolution as part of the creep process.

On the other hand, there are non-classical%
\footnote{By non-classical, we do not mean to imply that quantum mechanics is
used in the present paper.%
} models of multicomponent fluid systems in which interfaces between
phases are treated via the gradient thermodynamics approach\cite{Bloch1932,Ginzburg1950,Cahn58a}
also called the phase field method (see e.g.\ Ref.~\cite{Sekerka2011}
and references therein). The fluid theories include rigorous derivations
of the entropy production rate for the simultaneous processes of phase-field
evolution, heat conduction, diffusion and convective flows accompanied
by viscous dissipation. However, extensions of such theories to solid
materials are presently lacking. The existing phase-field models of
creep in solids\cite{Zhou2010} describe creep deformation though
a set of phase fields related to dislocations in specific slip systems.
Such theories reproduce creep-controlled structural evolution in multi-phase
materials without explicitly treating vacancies or the lattice. 

The goal of this paper is to develop a general irreversible thermodynamics
framework for the description of creep deformation in solid materials
by mechanisms involving site generation and annihilation and vacancy
diffusion. The proposed theory includes all three components (i)-(iii)
mentioned above. It can be viewed as a generalization of the non-classical
fluid theories\cite{Sekerka2011} to solid materials. Alternatively,
it can be considered as a generalization of classical solid-state
thermodynamics\cite{Larche73,Larche_Cahn_78,Larche1985,Mullins1985}
to non-classical, non-equilibrium solid systems with a non-conserved
lattice.

In Secs.~\ref{sec:Kinematics} and \ref{sec:Balance-equations} we
introduce the kinematic description of creep deformation and the balance
relations that will be used in the rest of the paper. Sect.~\ref{sec:Reversible-TD}
presents a thermodynamic treatment of a non-classical, non-hydrostatically
stressed multi-component solid phase. We derive gradient and time-dependent
forms of the first and second laws for reversible thermodynamic processes
in such a solid, along with a generalized form of the Gibbs-Duhem
equation. Before proceeding to irreversible thermodynamics, we derive
the conditions of full and constrained thermodynamic equilibria in
the solid. These conditions constitute a generalization of Larché
and Cahn\cite{Larche73,Larche_Cahn_78,Larche1985} to non-classical
solids with continuously distributed non-conserved sites. The entropy
production rate derived in Sect.~\ref{sec:Irreversible-TD} serves
as the starting point for the identification of the relevant fluxes
and forces and formulation of phenomenological relations between them.
We emphasize the importance of symmetry properties of the material,
formulate a set of phenomenological relations for isotropic materials,
and outline possible extensions to lower-symmetry systems by analyzing
the tensor character of the fluxes and forces. The volume and shear
components of the creep deformation rate\cite{Svoboda2006,Fischer2011}
emerge naturally from this analysis and are shown to be coupled to
different driving forces. To provide a simple illustration of how
the theory can be applied, we present a one-dimensional model of a
bicrystal with a grain boundary (GB) acting as a sink and source of
vacancies (Sect.~\ref{sec:Simple-model}). In the presence of vacancy
over-saturation or under an applied tensile stress, the kinetic equations
describe creep deformation of the sample accompanied by GB migration
and relative rigid translations of the grains. In Sect.~\ref{sec:Discussion}
we summarize the work and draw conclusions.

\section{Mass and site conservation laws and kinematics of deformation\label{sec:Kinematics}}

We consider a crystalline solid composed of $n$ substitutional chemical
species labeled $i$. The solid contains vacancies but not interstitials,
although this theory can be generalized to incorporate interstitials.
We assume that there are no chemical reactions among the species $i$.
The crystalline structure is assumed to have a Bravais lattice, i.e.,
primitive lattice with a single basis site. The solid is subject to
external potential forces such as gravitational or electric (when
the particles are electrically charged as in ionic solids). 

We start by formulating mass and particle conservation conditions
satisfied by our system. Some of them are specific to a solid solution
while others are equally valid for liquids or gases. The substitutional
lattice sites, referred to below as simply sites, are generally not
conserved. It is assumed, however, that we can still define a lattice
velocity field $\mathbf{v}_{L}(\mathbf{x},t)$. To this end, we assume
that the solid contains an imaginary network of sites which, on the
timescale of our observations, are not destroyed by the creep process.
These indestructible lattice sites will be called ``markers''.%
\footnote{The term ``marker'' may sound somewhat confusing because of the association
with the Kirkendall experiment\cite{Philibert} in which the markers
were inert foreign objects intentionally embedded in the lattice.
In our case the imaginary marker sites are physically identical to
other sites except for our knowledge that they will ``survive'' the
lattice site creation and annihilation during the creep deformation
process on a chosen timescale.%
} The lattice velocity $\mathbf{v}_{L}(\mathbf{x},t)$, also referred
to as the total lattice velocity, is defined as the velocity of a
marker occupying the location $\mathbf{x}$ (relative to a fixed laboratory
coordinate system) at a time $t$. We assume that the network of markers
is dense enough to treat $\mathbf{v}_{L}(\mathbf{x},t)$ as a continuous
function of coordinates. 

The number density $n_{s}(\mathbf{x},t)$ of the lattice sites per
unit volume satisfies the balance equation%
\footnote{We follow the convention\cite{Malvern69} that the dot between vectors
or tensor (e.g., $\mathbf{a}\cdot\mathbf{b}$) denotes their inner
product (contraction) while juxtaposition (e.g, $\mathbf{a}\mathbf{b}$)
their outer (dyadic) product. Two dots denote the double contractions
$\mathbf{a}\cdot\cdot\mathbf{b}=\textrm{Tr}(\mathbf{a}\cdot\mathbf{b})$
and $\mathbf{a}:\mathbf{b}=\mathbf{a}\cdot\cdot\mathbf{b^{T}}$, where
$\mathbf{a}$ and $\mathbf{b}$ are second-rank tensors and superscript
$T$ denotes transposition. The differentiation operator $\nabla$
is treated as a vector.%
}
\begin{equation}
\dfrac{\partial n_{s}}{\partial t}+\nabla\cdot\left(n_{s}\mathbf{v}_{L}\right)=r_{s},\label{eq:4}
\end{equation}
where $r_{s}(\mathbf{x},t)$ is the site generation rate (number of
sites per unit volume per unit time). The sign of $r_{s}$ is positive
for site generation and negative for annihilation.\cite{Stephenson1988}
This equation can be rewritten
\begin{equation}
\dfrac{d^{L}n_{s}}{dt}+n_{s}\nabla\cdot\mathbf{v}_{L}=r_{s},\label{eq:4-1}
\end{equation}
where the lattice material time derivative $d^{L}/dt$ is defined
by
\begin{equation}
\dfrac{d^{L}}{dt}\equiv\dfrac{\partial}{\partial t}+\mathbf{v}_{L}\mathbf{\cdot}\nabla.\label{eq:9}
\end{equation}

The number density $n_{i}$ of each material species $i$ obeys the
particle conservation law 
\begin{equation}
\dfrac{\partial n_{i}}{\partial t}+\nabla\cdot\left(n_{i}\mathbf{v}_{L}+\mathbf{J}_{i}^{L}\right)=0,\label{eq:5}
\end{equation}
or
\begin{equation}
\dfrac{d^{L}n_{i}}{dt}+n_{i}\nabla\cdot\mathbf{v}_{L}+\nabla\cdot\mathbf{J}_{i}^{L}=0,\label{eq:8}
\end{equation}
where $\mathbf{J}_{i}^{L}\equiv n_{i}(\mathbf{v}_{i}-\mathbf{v}_{L})$
is the diffusion flux of species $i$ relative to the lattice and
$\mathbf{v}_{i}$ is its observed velocity relative to the laboratory.

\begin{figure}
\noindent \begin{centering}
\includegraphics[scale=0.65]{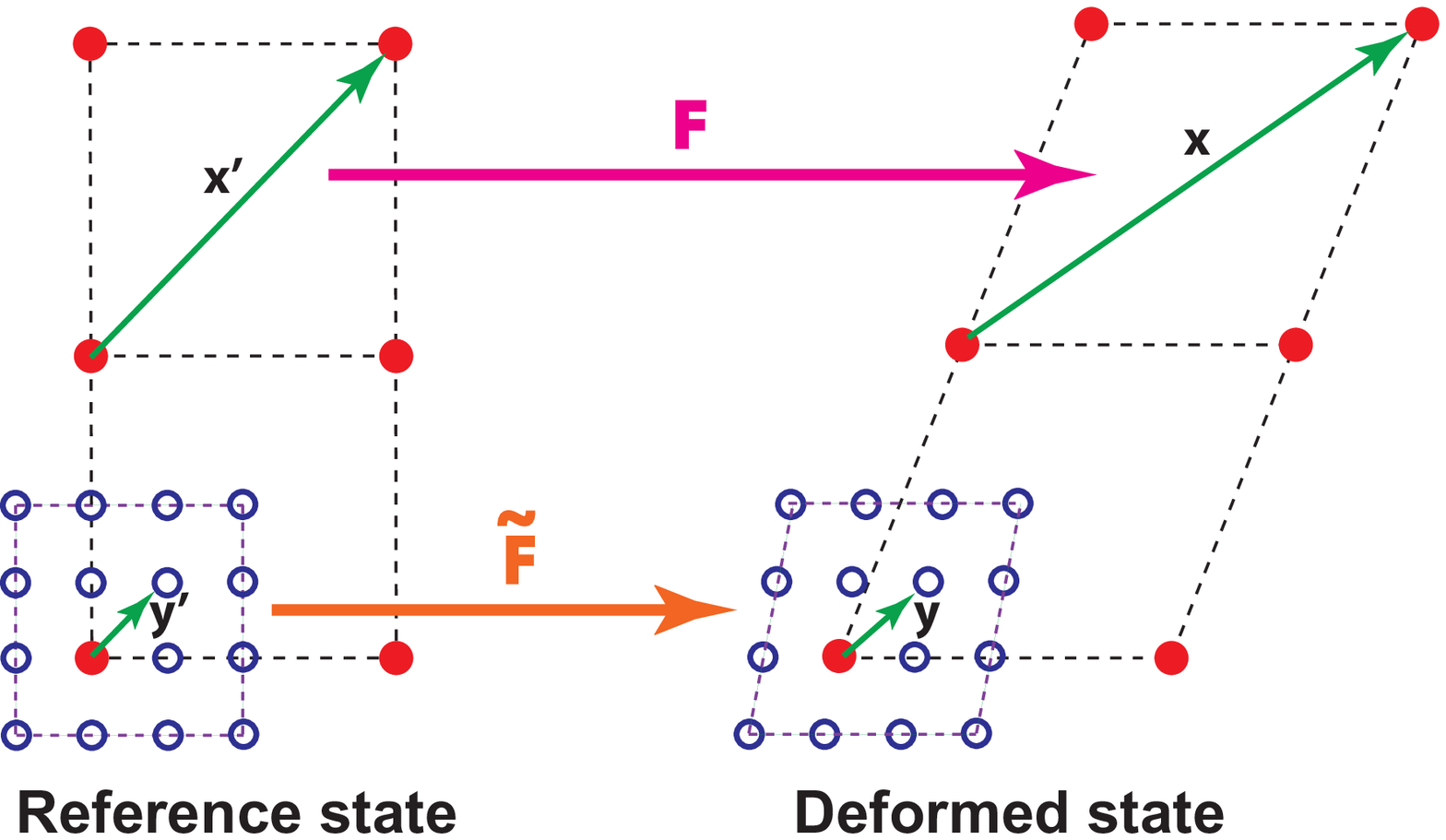}
\par\end{centering}

\caption{Dual-scale deformation of a solid material with site creation and
annihilation. The marker sites (filled circles) and regular lattice
sites (open circles) are connected by dashed lines to facilitate their
tracking during the deformation. The shape deformation gradient $\mathbf{F}$
is defined by the motion of the markers, whereas the local lattice
deformation gradient $\tilde{\mathbf{F}}$ is defined by mapping of
lattice sites in the vicinity of markers. Note that the deformation
of the network of markers is different from the local lattice distortion.\label{fig:Dual-scale-deformation}}
\end{figure}

Since the markers are conserved during the deformation process, they
can be used to define a deformation mapping $\mathbf{x}=\mathbf{x}(\mathbf{x}^{\prime},t)$
with respect to a chosen reference state, $\mathbf{x}^{\prime}$,
of the material (Fig.\ref{fig:Dual-scale-deformation}). This mapping
defines the shape (or total) deformation gradient\cite{Malvern69}
\begin{equation}
\mathbf{F}\equiv\left(\dfrac{\partial\mathbf{x}}{\partial\mathbf{x}^{\prime}}\right)_{t}\label{eq:6-1}
\end{equation}
and is related to the total lattice velocity (i.e., the velocity of
the marker network) by
\begin{equation}
\mathbf{v}_{L}=\left(\dfrac{\partial\mathbf{x}}{\partial t}\right)_{\mathbf{x}^{\prime}}.\label{eq:6-2}
\end{equation}

If the material is crystalline, then besides $\mathbf{F}$ we can
also define another lattice deformation gradient $\tilde{\mathbf{F}}$.
\cite{Bilby60a} To do so, we assume that for any lattice site we
can identify instantaneous directions of the crystallographic axes
in its vicinity. This allows us to establish a \emph{local} mapping
between lattice vectors, $\mathbf{y}$ and $\mathbf{y}^{\prime}$,
in the current and reference states, respectively (Fig.\ref{fig:Dual-scale-deformation}).%
\footnote{The lattice vector mapping can break down in core regions of lattice
defects. It is assumed that such regions comprise a negligibly small
fraction of the material and do not occur in the neighborhood of the
markers.%
} The deformation gradient defined by 
\begin{equation}
\tilde{\mathbf{F}}\equiv\left(\dfrac{\partial\mathbf{y}}{\partial\mathbf{y}^{\prime}}\right)_{\mathbf{x}^{\prime},t}\label{eq:6-3}
\end{equation}
represents local lattice distortions in the vicinity of a marker site
$\mathbf{x}^{\prime}$. It should be emphasized that this definition
of $\tilde{\mathbf{F}}$ does not imply conservation of sites in the
vicinity of the marker. With time, some of the sites my disappear,
but their locations are then filled by other sites resulting is a
self-reproduced local crystalline structure. This structure can be
identified at any instant by examining the \emph{current} atomic positions
around the marker and establishing their mapping on the reference
crystal structure. Since $\tilde{\mathbf{F}}$ is defined in a \emph{small}
vicinity of every marker site $\mathbf{x}^{\prime}$, we assume that
it is independent of $\mathbf{y}^{\prime}$ and is a continuous function
of $\mathbf{x}^{\prime}$, i.e., $\tilde{\mathbf{F}}=\tilde{\mathbf{F}}(\mathbf{x}^{\prime},t)$.%
\footnote{The ability to describe lattice deformations by a single deformation
gradient $\tilde{\mathbf{F}}$ relies on the assumption of a Bravais
lattice of the crystal structure. Non-Bravais structures would require
additional variables describing internal strains of the unit cell. %
} 

Generally, $\mathbf{F}$ and $\tilde{\mathbf{F}}$ are two different
tensors. In particular, the derivative 
\begin{equation}
\mathbf{\tilde{v}}_{L}\equiv\left(\dfrac{\partial\mathbf{y}}{\partial t}\right)_{\mathbf{x}^{\prime},\mathbf{y}^{\prime}}\label{eq:6-4}
\end{equation}
defines the local lattice velocity $\mathbf{\tilde{v}}_{L}$ due to
elastic deformation, thermal expansion and compositional strains.
This velocity is generally different from the marker network velocity
$\mathbf{v}_{L}$. The latter incorporates the same deformation effects
as $\mathbf{\tilde{v}}_{L}$ but additionally includes the permanent
deformation due to site generation and annihilation.

Thus, we introduce \emph{two} \emph{different} deformation gradients
between the \emph{same} pair of deformed and reference states of the
material: the shape deformation gradient $\mathbf{F}$ defined by
the marker-to-marker mapping, and the lattice deformation gradient
$\tilde{\mathbf{F}}$ defined by local lattice mapping in the vicinity
of every marker. The lattice site generation and annihilation during
the creep process produces permanent deformation leading to deviations
of $\mathbf{F}$ from $\tilde{\mathbf{F}}$. Experimentally, information
about $\tilde{\mathbf{F}}$ could be obtained by X-ray diffraction
measurements whereas $\mathbf{F}$ could be simultaneously measured
by dilatometry. This type of measurements were used by Simmons and
Balluffi\cite{Simmons1960,Simmons1960b} to determine the equilibrium
vacancy concentration in metals. 

This dual description of deformation is central to our theory and
will be employed for the calculations of the entropy production rate
in the materials and other kinetic characteristics of diffusional
creep.

There is an important kinematic relation between the two velocities
$\mathbf{v}_{L}$ and $\mathbf{\tilde{v}}_{L}$, on one hand, and
the lattice site production rate $r_{s}$ on the other. To derive
it, return to the site balance Eq.(\ref{eq:4-1}). This equation can
be rewritten in the form
\begin{equation}
\dfrac{d^{L}n_{s}}{dt}\equiv\left(\dfrac{\partial n_{s}}{\partial t}\right)_{\mathbf{x}^{\prime}}=r_{s}-n_{s}\nabla_{\mathbf{x}}\cdot\mathbf{v}_{L}.\label{eq:6-7}
\end{equation}
On the other hand, the site density can be expressed as
\begin{equation}
n_{s}=\dfrac{n_{s}^{\prime}}{\tilde{G}},\label{eq:6-5}
\end{equation}
where $\tilde{G}\equiv\det\tilde{\mathbf{F}}$ and $n_{s}^{\prime}$
is the lattice site density in the reference state, assumed to be
a constant. Using the Jacobi identity\cite{Malvern69} it can be shown
that
\begin{equation}
\dfrac{d^{L}\tilde{G}}{dt}\equiv\left(\dfrac{\partial\tilde{G}}{\partial t}\right)_{\mathbf{x}^{\prime}}=\tilde{G}\nabla_{\mathbf{y}}\cdot\mathbf{\tilde{v}}_{L},\label{eq:6-6}
\end{equation}
where we used the local lattice mapping $\mathbf{y}=\mathbf{y}(\mathbf{y}^{\prime},\mathbf{x}^{\prime},t)$
considering the marker position $\mathbf{x}^{\prime}$ as a parameter.
Applying this relation to Eq.(\ref{eq:6-5}) we have
\begin{equation}
\dfrac{d^{L}n_{s}}{dt}\equiv\left(\dfrac{\partial n_{s}}{\partial t}\right)_{\mathbf{x}^{\prime}}=n_{s}^{\prime}\left(\dfrac{\partial(1/\tilde{G})}{\partial t}\right)_{\mathbf{x}^{\prime}}=-n_{s}\nabla_{\mathbf{y}}\cdot\mathbf{\tilde{v}}_{L}.\label{eq:6-8}
\end{equation}

There is a subtle difference between Eqs.(\ref{eq:6-7}) and (\ref{eq:6-8}).
In Eq.(\ref{eq:6-7}), $n_{s}$ is the coarse-grained site density
averaged over a volume containing a group of neighboring markers,
whereas in Eq.(\ref{eq:6-8}) $n_{s}$ is a more detailed function
of coordinates near a particular marker $\mathbf{x}^{\prime}$. Assuming
that $n_{s}$ is a slowly varying function of coordinates on the scale
of inter-marker distances, we treat both densities as equal and their
time derivatives in Eqs.(\ref{eq:6-7}) and (\ref{eq:6-8}) as identical.
This immediately gives
\begin{equation}
\nabla\cdot\mathbf{v}_{L}-\nabla\cdot\mathbf{\tilde{v}}_{L}=\dfrac{r_{s}}{n_{s}}.\label{eq:6-9}
\end{equation}
We dropped the subscripts of the divergence symbols, but it should
be remembered that the divergence of $\mathbf{\tilde{v}}_{L}$ is
taken locally whereas the divergence of $\mathbf{v}_{L}$ is coarse-grained
over a volume containing multiple markers. 

Eq.(\ref{eq:6-9}) reflects the fact that the site generation causes
deviations of the total velocity divergence $\nabla\cdot\mathbf{v}_{L}$
from the local velocity divergence $\nabla\cdot\mathbf{\tilde{v}}_{L}$
arising solely from lattice distortions. In the absence of site generation,
the two velocity fields are identical and Eq.(\ref{eq:6-9}) correctly
predicts $r_{s}=0$. We will show later that $r_{s}$ is the trace
of a more general tensor representing a more complete view of the
permanent deformation caused by site generation and annihilation.

\section{Balance equations\label{sec:Balance-equations}}

In this Section we summarize the momentum, energy and entropy balance
relations that will be used in this work and discuss the assumptions
and approximations underlying these relations.

\subsection{Momentum balance}

For our multicomponent system, it is necessary to derive a consistent
momentum balance equation. The standard momentum equation for a single
component solid, such as treated by Malvern,\cite{Malvern69} is no
longer applicable because of the momentum carried by multicomponent
diffusion. As shown in Appendix A, the correct momentum equation is
\begin{equation}
\dfrac{\partial}{\partial t}\left(\rho\mathbf{v}\right)+\nabla\cdot\left(\sum_{i=1}^{n}m_{i}n_{i}\mathbf{v}_{i}\mathbf{v}_{i}\right)=\mathbf{b}+\nabla\cdot\boldsymbol{\sigma},\label{eq:10}
\end{equation}
where $\mathbf{v}$ is the barycentric velocity, $\rho$ is the material
density (mass per unit volume), $\mathbf{b}=\sum_{i=1}^{n}n_{i}\mathbf{b}_{i}$
is the external force per unit volume, and $\nabla\cdot\boldsymbol{\sigma}$
is the force exerted by the stress $\boldsymbol{\sigma}$ per unit
volume of the material. We assume that the external fields are conservative,
so that the force per particle 
\begin{equation}
\mathbf{b}_{i}=-\nabla\psi_{i},\label{eq:12-1}
\end{equation}
where $\psi_{i}$ are species-specific potential functions.

Eq.(\ref{eq:10}) can be rewritten with respect to the lattice (see
Appendix A)
\begin{equation}
\rho\dfrac{d^{L}\mathbf{v}_{L}}{dt}=\mathbf{b}+\nabla\cdot\left(\boldsymbol{\sigma}-\mathbf{M}\right)-\dfrac{d^{L}\boldsymbol{q}}{dt}-\boldsymbol{q}\nabla\cdot\mathbf{v}_{L}-\boldsymbol{q}\cdot\nabla\mathbf{v}_{L},\label{eq:12-2}
\end{equation}
where tensor $\mathbf{M}$ is given by 
\begin{equation}
\mathbf{M}\equiv\sum_{i=1}^{n}\dfrac{m_{i}}{n_{i}}\mathbf{J}_{i}^{L}\mathbf{J}_{i}^{L}\label{eq:13-2}
\end{equation}
and vector
\begin{equation}
\boldsymbol{q}\equiv\sum_{i=1}^{n}m_{i}\mathbf{J}_{i}^{L}=\rho\left(\mathbf{v}-\mathbf{v}_{L}\right)\label{eq:6-1-1}
\end{equation}
is the momentum density carried by the local center of mass relative
to the lattice. Here, $m_{i}$ is the mass of particles of species
$i$. The derivative $d^{L}\boldsymbol{q}/dt$ is the inertia force
which arises due to the fact that the lattice and barycentric references
are both non-inertial.

\subsection{Energy balance\label{sub:Energy-balance}}

The total energy $e$ of the material per unit volume can be expressed
\begin{equation}
e=K+\psi+u.\label{eq:28}
\end{equation}
where
\begin{equation}
K=\dfrac{1}{2}\sum_{i=1}^{n}m_{i}n_{i}|\mathbf{v}_{i}|^{2}\label{eq:38}
\end{equation}
is the macroscopic kinetic energy of the particles per unit volume,
\begin{equation}
\psi\equiv\sum_{i=1}^{n}n_{i}\psi_{i}\label{eq:11-1}
\end{equation}
is potential energy in external fields per unit volume, and the remaining
term $u$ is identified with internal energy per unit volume. The
latter includes the energy of interactions between the particles and
the kinetic energy of their microscopic motion (e.g., lattice vibrations,
molecular rotations, etc.), but excludes the macroscopic kinetic energy
due to diffusion. The internal energy can be shown to satisfy the
following balance equation with respect to the lattice (see Appendix
A):
\begin{eqnarray}
\dfrac{d^{L}u}{dt}+u\nabla\cdot\mathbf{v}_{L} & = & -\nabla\cdot\mathbf{J}_{u}^{L}+\sum_{i=1}^{n}\mathbf{b}_{i}\cdot\mathbf{J}_{i}^{L}+\left(\boldsymbol{\sigma}-\mathbf{M}\right):\nabla\mathbf{v}_{L}\nonumber \\
 &  & -\sum_{i=1}^{n}\left\{ \nabla\left[\dfrac{m_{i}}{2n_{i}^{2}}\left(\mathbf{J}_{i}^{L}\cdot\mathbf{J}_{i}^{L}\right)\right]+m_{i}\dfrac{d^{L}\mathbf{v}_{i}}{dt}\right\} \cdot\mathbf{J}_{i}^{L},\label{eq:13}
\end{eqnarray}
where $\mathbf{J}_{u}^{L}$ is the internal energy flux relative to
the lattice.

Equation (\ref{eq:13}) is valid for all, not necessarily reversible,
process and expresses the first law of thermodynamics stating that
the change in internal energy equals the work done on the system less
the energy dissipated through its boundaries. As with the momentum
balance relation (\ref{eq:12-2}), Eq.(\ref{eq:13}) is exact: it
represents the internal energy balance without any approximations
or assumptions other than the conservation of energy and the total
energy ansatz (\ref{eq:28}).

We will also need the potential energy balance relation, 
\begin{equation}
\dfrac{d^{L}\psi}{dt}+\psi\nabla\cdot\mathbf{v}_{L}=-\sum_{i=1}^{n}\mathbf{b}_{i}\cdot\mathbf{J}_{i}^{L}-\mathbf{b}\cdot\mathbf{v}_{L}-\nabla\cdot\left(\sum_{i=1}^{n}\psi_{i}\mathbf{J}_{i}^{L}\right),\label{eq:13-3}
\end{equation}
where the last term represents the divergence of the diffusive flux
of potential energy. This relation is also exact.

\subsection{Entropy balance\label{sub:Entropy-balance}}

The entropy balance is postulated in the form
\begin{equation}
\dfrac{d^{L}s}{dt}+s\nabla\cdot\mathbf{v}_{L}+\nabla\cdot\mathbf{J}_{s}^{L}=\dot{s},\label{eq:15}
\end{equation}
where $s$ is entropy per unit volume, $\mathbf{J}_{s}^{L}$ is the
entropy flux carried by the conduction of heat relative to the lattice,
and $\dot{s}$ is the entropy production rate due to irreversible
processes. 

The goal of the subsequent development will be to compute $\dot{s}$.
The common approach\cite{De-Groot1984} to achieving this goal is
to calculate the entropy rate $(d^{L}s/dt+s\nabla\cdot\mathbf{v}_{L})$
and then rearrange the terms to form the divergence of fluxes that
can be identified with $-\nabla\cdot\mathbf{J}_{s}^{L}$. The remaining
terms are then identified with $\dot{s}$. We will follow this route
to derive $\dot{s}$ for a solid material containing non-conserved
lattice sites.

\section{Local reversible thermodynamics\label{sec:Reversible-TD}}

\subsection{The local equilibrium postulate}

It is assumed that, although the entire solid material can be away
from equilibrium, its \emph{local} internal energy, entropy and other
thermodynamic variables are related to each other via a fundamental
equation of state describing \emph{reversible} processes. ``Local''
means here that this equation is followed only by subsystems of the
entire system that are small enough to reach thermodynamic equilibrium
before the entire system does, yet large enough to apply the full
formalism of thermodynamics. The locally equilibrium subsystems need
not be uniform and can be treated using the formalism of gradient
thermodynamics.\cite{Bloch1932,Ginzburg1950,Cahn58a}

Relative to the moving lattice, the fundamental equation is postulated
in the functional form:
\begin{equation}
u=u\left(s,\{n_{i}\},\{\varphi_{k}\},\{\nabla n_{i}\},\{\nabla\varphi_{k}\},\tilde{\mathbf{F}}\right).\label{eq:16}
\end{equation}
Here, $\varphi_{k}$ ($k=1,...,K$) are non-conserved phase fields,
$\nabla n_{i}$ and $\nabla\varphi_{k}$ are respective gradients,
and $\tilde{\mathbf{F}}$ is the lattice deformation gradient relative
to a chosen reference state (Sec.~\ref{sec:Kinematics}). The phase
fields $\varphi_{k}$ can represent different thermodynamic phases
of the material or be associated with different lattice orientations
(grains) in a single-phase polycrystalline material. The gradients
$\nabla\varphi_{k}$ and $\nabla n_{i}$ are usually negligibly small
inside the bulk phases or grains but are important in the description
of inter-phase interfaces and GBs. The material regions whose thermodynamic
description requires the gradients\cite{Bloch1932,Ginzburg1950,Cahn58a}
are referred to as ``non-classical'' as opposed to ``classical'' regions
which can be treated within the standard thermodynamics\cite{Willard_Gibbs}
of homogeneous phases. Since $u$ is a scalar while the gradients
are vectors and $\tilde{\mathbf{F}}$ is a tensor, it is assumed that
Eq.(\ref{eq:16}) satisfies the required invariance under rotations
of the coordinate system. 

When Eq.(\ref{eq:16}) is applied to different locations in the solid,
it is assumed that the reference state used to describe the lattice
deformation is the same for every location and is fixed once and for
all. For example, for a cubic crystal the reference state can be a
perfectly cubic unit cell with a given (e.g., stress-free) lattice
constant. This explains why properties of the reference state, such
as the reference volume per site, are not listed among the variables
of $u$.

\subsection{The first and second laws of thermodynamics for local reversible
processes}

To derive the differential form of Eq.(\ref{eq:16}), let us first
consider a uniform region containing a fixed number of lattice sites.
Suppose for the moment that the phase fields $\varphi_{k}$ are not
included. The standard differential form of the fundamental equation
for such a region is 
\begin{equation}
dU=TdS+\sum_{i=1}^{n}M_{i}dN_{i}+V\left(\tilde{\mathbf{F}}^{-1}\cdot\tilde{\mathbf{\boldsymbol{\sigma}}}\right)\cdot\cdot d\tilde{\mathbf{F}}.\label{eq:16-1}
\end{equation}
Here $U=uV$, $S=sV$ and $N_{i}=n_{i}V$ are the total internal energy,
entropy and numbers of particles of the chemical components inside
the region, $V$ is its volume, $T\equiv\partial U/\partial S$ is
temperature and $\tilde{\mathbf{F}}^{-1}$ is the inverse of $\tilde{\mathbf{F}}$.
The tensor $\tilde{\mathbf{\boldsymbol{\sigma}}}$ is formally \emph{defined}
through the derivative $\partial U/\partial\mathbf{\tilde{\mathbf{F}}}$,
\begin{equation}
\tilde{\mathbf{\boldsymbol{\sigma}}}\equiv\dfrac{1}{V}\mathbf{\tilde{\mathbf{F}}}\cdot\dfrac{\partial U}{\partial\mathbf{\tilde{\mathbf{F}}}},\label{eq:16-2}
\end{equation}
and has the meaning of the equilibrium Cauchy stress in a uniform
lattice. As will be discussed later, $\tilde{\mathbf{\boldsymbol{\sigma}}}$
it is generally different from the actual stress tensor $\boldsymbol{\sigma}$
in a non-uniform and/or non-equilibrium material. The obvious motivation
behind the definition (\ref{eq:16-2}) is the standard form $V^{\prime}\mathbf{P}\cdot\cdot d\tilde{\mathbf{F}}$
of the mechanical work term in continuum mechanics\cite{Malvern69},
$V^{\prime}$ being the reference volume of the region and $\mathbf{P}=\tilde{J}\tilde{\mathbf{F}}^{-1}\cdot\tilde{\mathbf{\boldsymbol{\sigma}}}$
the first Piola-Kirchhoff stress tensor. Finally, the derivative $M_{i}\equiv\partial U/\partial N_{i}$
has the meaning of the diffusion potential of species $i$ relative
to vacancies if the latter are treated as massless species. If only
the material particles are treated as species, $M_{i}$ can be considered
as simply the chemical potential of species $i$. As discussed in
the literature\cite{Larche1985,Mullins1985}, both interpretations
of $M_{i}$ are equally legitimate and give the same results for all
physically observable quantities.

Eq.(\ref{eq:16-1}) can be rewritten in terms of the volume densities
$u$, $s$ and $n_{i}$ : 
\begin{equation}
du=Tds+\sum_{i=1}^{n}M_{i}dn_{i}+\left(\tilde{\mathbf{F}}^{-1}\cdot\tilde{\mathbf{\boldsymbol{\sigma}}}\right)\cdot\cdot d\tilde{\mathbf{F}}-\left(u-Ts-\sum_{i=1}^{n}M_{i}n_{i}\right)\dfrac{dV}{V}.\label{eq:17}
\end{equation}
Using the identity\cite{Malvern69} 
\begin{equation}
\dfrac{dV}{V}=\tilde{\mathbf{F}}^{-1}\cdot\cdot d\tilde{\mathbf{F}}\label{eq:17-1}
\end{equation}
we obtain
\begin{equation}
du=Tds+\sum_{i=1}^{n}M_{i}dn_{i}+\left(\tilde{\mathbf{F}}^{-1}\cdot\left(\tilde{\mathbf{\boldsymbol{\sigma}}}-\omega\mathbf{I}\right)\right)\cdot\cdot d\tilde{\mathbf{F}},\label{eq:17-2}
\end{equation}
where $\mathbf{I}$ is the second rank unit tensor and
\begin{equation}
\omega\equiv u-Ts-\sum_{i=1}^{n}M_{i}n_{i}\label{eq:30-1}
\end{equation}
is the grand-canonical potential per unit volume. 

Eq.(\ref{eq:17-2}) is the differential form of Eq.(\ref{eq:16})
for the particular case of a uniform material without phase fields.
In the presence of phase fields and the gradients $\nabla\varphi_{k}$
and $\nabla n_{i}$, this equation becomes
\begin{eqnarray}
du & = & Tds+\sum_{i=1}^{n}M_{i}dn_{i}+\sum_{k=1}^{K}\dfrac{\partial u}{\partial\varphi_{k}}d\varphi_{k}+\sum_{i=1}^{n}\dfrac{\partial u}{\partial\nabla n_{i}}\cdot d\nabla n_{i}+\sum_{k=1}^{K}\dfrac{\partial u}{\partial\nabla\varphi_{k}}\cdot d\nabla\varphi_{k}\nonumber \\
 & + & \left(\tilde{\mathbf{F}}^{-1}\cdot\left(\tilde{\mathbf{\boldsymbol{\sigma}}}-\omega\mathbf{I}\right)\right)\cdot\cdot d\tilde{\mathbf{F}}.\label{eq:17-3}
\end{eqnarray}
Note that $\omega$ appearing in the last term is now a non-classical
quantity as it depends on the gradients through $u$ {[}cf.~Eq.(\ref{eq:16}){]}.

The gradient terms in Eq.(\ref{eq:17-3}) can be rearranged using
the identities
\begin{eqnarray}
\sum_{i=1}^{n}\dfrac{\partial u}{\partial\nabla n_{i}}\cdot d\nabla n_{i} & = & \sum_{i=1}^{n}\nabla\cdot\left(\dfrac{\partial u}{\partial\nabla n_{i}}dn_{i}\right)-\sum_{i=1}^{n}\left(\nabla\cdot\dfrac{\partial u}{\partial\nabla n_{i}}\right)dn_{i}\label{eq:17-4}\\
 & + & \sum_{i=1}^{n}\dfrac{\partial u}{\partial\nabla n_{i}}\cdot\left(d\nabla n_{i}-\nabla dn_{i}\right),\nonumber 
\end{eqnarray}
\begin{eqnarray}
\sum_{k=1}^{K}\dfrac{\partial u}{\partial\nabla\varphi_{k}}\cdot d\nabla\varphi_{k} & = & \sum_{k=1}^{K}\nabla\cdot\left(\dfrac{\partial u}{\partial\nabla\varphi_{k}}d\varphi_{k}\right)-\sum_{k=1}^{K}\left(\nabla\cdot\dfrac{\partial u}{\partial\nabla\varphi_{k}}\right)d\varphi_{k}\nonumber \\
 & + & \sum_{k=1}^{K}\dfrac{\partial u}{\partial\nabla\varphi_{k}}\cdot\left(d\nabla\varphi_{k}-\nabla d\varphi_{k}\right),\label{eq:17-5}
\end{eqnarray}
where we recognize that the operations $d$ and $\nabla$ may not
commute. Equation (\ref{eq:17-3}) finally becomes
\begin{eqnarray}
du & = & Tds+\sum_{i=1}^{n}M_{i}^{*}dn_{i}+\sum_{k=1}^{K}\Phi_{k}^{*}d\varphi_{k}+\nabla\cdot\left(\sum_{i=1}^{n}\dfrac{\partial u}{\partial\nabla n_{i}}dn_{i}+\sum_{k=1}^{K}\dfrac{\partial u}{\partial\nabla\varphi_{k}}d\varphi_{k}\right)\nonumber \\
 & + & \sum_{i=1}^{n}\dfrac{\partial u}{\partial\nabla n_{i}}\cdot\left(d\nabla n_{i}-\nabla dn_{i}\right)+\sum_{k=1}^{K}\dfrac{\partial u}{\partial\nabla\varphi_{k}}\cdot\left(d\nabla\varphi_{k}-\nabla d\varphi_{k}\right)\nonumber \\
 & + & \left(\tilde{\mathbf{F}}^{-1}\cdot\left(\tilde{\mathbf{\boldsymbol{\sigma}}}-\omega\mathbf{I}\right)\right)\cdot\cdot d\tilde{\mathbf{F}},\label{eq:17-3-1}
\end{eqnarray}
where%
\footnote{For clarity, some of the non-classical quantities will be designated
by an asterisk. %
}
\begin{equation}
M_{i}^{*}\equiv\dfrac{\partial u}{\partial n_{i}}-\nabla\cdot\dfrac{\partial u}{\partial\nabla n_{i}}=M_{i}-\nabla\cdot\dfrac{\partial u}{\partial\nabla n_{i}}\label{eq:17-3-3}
\end{equation}
is the non-classical diffusion potential and 
\begin{equation}
\Phi_{k}^{*}\equiv\dfrac{\partial u}{\partial\varphi_{k}}-\nabla\cdot\dfrac{\partial u}{\partial\nabla\varphi_{k}}.\label{eq:17-3-2}
\end{equation}
Note that $M_{i}^{*}$ and $\Phi_{k}^{*}$ are variational derivatives\cite{Gelfand-Fomin}
of the internal energy with respect to the concentrations $n_{i}$
and phase fields $\varphi_{k}$, respectively.

The obtained Eq.(\ref{eq:17-3-1}) is a formulation of the first and
second laws of thermodynamics for local reversible processes in a
lattice obeying the fundamental equation (\ref{eq:16}). It will serve
as the starting point for several derivations performed below.

\subsection{Generalized Gibbs-Duhem equation\label{sub:Gibbs-Duhem}}

By applying a partial Legendre transformation \cite{Gelfand-Fomin}
with respect to $s$ and $n_{i}$, Eq.(\ref{eq:17-3-1}) can be transformed
to 
\begin{eqnarray}
 &  & sdT+\sum_{i=1}^{n}n_{i}dM_{i}^{*}+d\omega-\sum_{k=1}^{K}\Phi_{k}^{*}d\varphi_{k}+\sum_{i=1}^{n}d\left(n_{i}\nabla\cdot\dfrac{\partial u}{\partial\nabla n_{i}}\right)\nonumber \\
 &  & -\nabla\cdot\left(\sum_{i=1}^{n}\dfrac{\partial u}{\partial\nabla n_{i}}dn_{i}+\sum_{k=1}^{K}\dfrac{\partial u}{\partial\nabla\varphi_{k}}d\varphi_{k}\right)\nonumber \\
 &  & -\sum_{i=1}^{n}\dfrac{\partial u}{\partial\nabla n_{i}}\cdot\left(d\nabla n_{i}-\nabla dn_{i}\right)-\sum_{k=1}^{K}\dfrac{\partial u}{\partial\nabla\varphi_{k}}\cdot\left(d\nabla\varphi_{k}-\nabla d\varphi_{k}\right)\nonumber \\
 &  & -\left(\tilde{\mathbf{F}}^{-1}\cdot\left(\tilde{\mathbf{\boldsymbol{\sigma}}}-\omega\mathbf{I}\right)\right)\cdot\cdot d\tilde{\mathbf{F}}=0.\label{eq:17-3-1-1}
\end{eqnarray}
This equation can be viewed as a generalization of the Gibbs-Duhem
equation\cite{Willard_Gibbs} to a non-classical solid subject to
non-hydrostatic mechanical stresses. In the particular case of a hydrostatically
stressed classical (no gradients) solid we have $\tilde{\mathbf{\boldsymbol{\sigma}}}=-p\mathbf{I}$,
where $p=-\omega$ is the equilibrium hydrostatic pressure, and Eq.(\ref{eq:17-3-1-1})
reduces to the standard Gibbs-Duhem equation\cite{Willard_Gibbs}
\begin{equation}
sdT+\sum_{i=1}^{n}n_{i}dM_{i}-dp=0.\label{eq:202-1}
\end{equation}

As an application of Eq.(\ref{eq:17-3-1-1}), suppose the differentials
represent infinitesimal differences between the values of properties
at two nearby points $\mathbf{x}$ and $\mathbf{x}+d\mathbf{x}$ at
a fixed moment of time. Then $dT=\nabla T\cdot d\mathbf{x}$, $dM_{i}^{*}=\nabla M_{i}^{*}\cdot d\mathbf{x}$,
and similarly for all other terms. In this particular case the operators
$d$ and $\nabla$ commute, $\nabla d=d\nabla=d\mathbf{x}\cdot\nabla\nabla$,
and both sums in the third line of Eq.(\ref{eq:17-3-1-1}) vanish.
The remaining terms contain the common factor $d\mathbf{x}$ which
cancels, giving
\begin{equation}
s\nabla T+\sum_{i=1}^{n}n_{i}\nabla M_{i}^{*}-\sum_{k=1}^{K}\Phi_{k}^{*}\nabla\varphi_{k}+\nabla\cdot\left(\mathbf{A}^{*}+\omega\mathbf{I}\right)-\left(\tilde{\mathbf{F}}^{-1}\cdot\left(\tilde{\mathbf{\boldsymbol{\sigma}}}-\omega\mathbf{I}\right)\right)\cdot\cdot\left(\tilde{\mathbf{F}}\overleftarrow{\nabla}\right)=0,\label{eq:17-6}
\end{equation}
where 
\begin{equation}
\mathbf{A^{*}}\equiv\left(\sum_{i=1}^{n}n_{i}\nabla\cdot\dfrac{\partial u}{\partial\nabla n_{i}}\right)\mathbf{I}-\sum_{i=1}^{n}\dfrac{\partial u}{\partial\nabla n_{i}}\nabla n_{i}-\sum_{k=1}^{K}\dfrac{\partial u}{\partial\nabla\varphi_{k}}\nabla\varphi_{k}\label{eq:23-1-1}
\end{equation}
is a purely non-classical second-rank tensor. 

Equation (\ref{eq:17-6}) is a gradient form of the generalized Gibbs-Duhem
equation (\ref{eq:17-3-1-1}). As will be shown later (Sec.~\ref{sec:Equilibrium-state}),
when the material reaches full thermodynamic equilibrium (including
equilibrium with respect to site generation and annihilation), the
first, third and last terms in Eq.(\ref{eq:17-6}) vanish while the
non-classical chemical potentials satisfy the condition $\nabla M_{i}^{*}-\mathbf{b}_{i}=\mathbf{0}.$
It follows that under the full equilibrium conditions
\begin{equation}
\nabla\cdot\left(\mathbf{A}^{*}+\omega\mathbf{I}\right)+\mathbf{b}=\mathbf{0}.\label{eq:18-1}
\end{equation}
Thus, in the absence of external fields, tensor $\left(\mathbf{A}^{*}+\omega\mathbf{I}\right)$
is divergence-free. In a one-dimensional system this means conservation
of the quantity $(A_{11}^{*}+\omega)$. 

The divergence-free character of $\left(\mathbf{A}^{*}+\omega\mathbf{I}\right)$
in the absence of external fields originates from the property of
the fundamental equation (\ref{eq:16}) that the internal energy does
not depend explicitly on the position vector $\mathbf{x}$. If it
did, an additional term $\partial u/\partial\mathbf{x}$ would appear
in Eq.(\ref{eq:17-3}) and eventually propagate to Eq.(\ref{eq:18-1}),
so that the divergence of $\left(\mathbf{A}^{*}+\omega\mathbf{I}\right)$
would no longer be zero. The mathematical procedure that produced
the divergence term in Eqs.(\ref{eq:17-6}) and (\ref{eq:18-1}) is
essentially equivalent to a derivation of Noether's theorem \cite{Gelfand-Fomin}
for a system with continuous translational symmetry. Applied fields
$\mathbf{b}_{i}$ obviously destroy this symmetry and lead to a nonzero
divergence of $\left(\mathbf{A}^{*}+\omega\mathbf{I}\right)$ as indicated
in Eq.(\ref{eq:18-1}).

\subsection{Time-dependent form of the first and second laws\label{sub:T-dependent-form}}

Returning to the general Eq.(\ref{eq:17-3-1}), we now consider the
case where the differentials represent changes in time. Because the
internal energy $u$ has been defined relative to the stationary lattice,
its time evolution must be described by the lattice material derivative
$d^{L}u/dt$ defined by Eq.(\ref{eq:8}). We will therefore interpret
all differentials $d$ in Eq.(\ref{eq:17-3-1}) as $d^{L}/dt$.

The operators $d^{L}/dt$ and $\nabla$ do not commute, but it can
be shown that%
\footnote{This follows from the definition of the material time derivative in
Eq.(\ref{eq:9}), the commutativity of $\partial/\partial t$ and
$\nabla$ and the vector identity $\nabla(\mathbf{v}_{L}\cdot\nabla)=\mathbf{v}_{L}\cdot\nabla\nabla+\nabla\mathbf{v}_{L}\cdot\nabla$.%
} 
\begin{equation}
\dfrac{d^{L}}{dt}\nabla-\nabla\dfrac{d^{L}}{dt}=-\nabla\mathbf{v}_{L}\cdot\nabla.\label{eq:23}
\end{equation}
As a result, the second line in Eq.(\ref{eq:17-3-1}) becomes
\[
-\sum_{i=1}^{n}\dfrac{\partial u}{\partial\nabla n_{i}}\cdot\nabla\mathbf{v}_{L}\cdot\nabla n_{i}-\sum_{k=1}^{K}\dfrac{\partial u}{\partial\nabla\varphi_{k}}\cdot\nabla\mathbf{v}_{L}\cdot\nabla\varphi_{k}
\]
and can be simplified to
\[
-\left(\sum_{i=1}^{n}\dfrac{\partial u}{\partial\nabla n_{i}}\nabla n_{i}+\sum_{k=1}^{K}\dfrac{\partial u}{\partial\nabla\varphi_{k}}\nabla\varphi_{k}\right):\nabla\mathbf{v}_{L}.
\]
The last term in Eq.(\ref{eq:17-3-1}) can be transformed to
\[
\left(\tilde{\mathbf{F}}^{-1}\cdot\left(\tilde{\mathbf{\boldsymbol{\sigma}}}-\omega\mathbf{I}\right)\right)\cdot\cdot\dfrac{d^{L}\tilde{\mathbf{F}}}{dt}=\left(\tilde{\mathbf{\boldsymbol{\sigma}}}-\omega\mathbf{I}\right):\nabla\mathbf{\tilde{v}}_{L},
\]
where we used the identity%
\footnote{Indeed, using the lattice mapping $\mathbf{y}(\mathbf{x}^{\prime},\mathbf{y}^{\prime},t)$
and Eqs.(\ref{eq:6-3}) and (\ref{eq:6-4}) we have $\dfrac{d^{L}\tilde{\mathbf{F}}}{dt}=\left(\dfrac{\partial\tilde{\mathbf{F}}}{\partial t}\right)_{\mathbf{x}^{\prime},\mathbf{y}^{\prime}}=\dfrac{\partial}{\partial t}\left(\left(\dfrac{\partial\mathbf{y}}{\partial\mathbf{y}^{\prime}}\right)_{\mathbf{x}^{\prime},t}\right)_{\mathbf{x}^{\prime},\mathbf{y}^{\prime}}=\dfrac{\partial}{\partial\mathbf{y}^{\prime}}\left(\left(\dfrac{\partial\mathbf{y}}{\partial t}\right)_{\mathbf{x}^{\prime},\mathbf{y}^{\prime}}\right)_{\mathbf{x}^{\prime},t}=\left(\dfrac{\partial\mathbf{\tilde{v}}_{L}}{\partial\mathbf{y}^{\prime}}\right)_{\mathbf{x}^{\prime},t}=\left(\dfrac{\partial\mathbf{\tilde{v}}_{L}}{\partial\mathbf{y}}\right)_{\mathbf{x}^{\prime},t}\left(\dfrac{\partial\mathbf{y}}{\partial\mathbf{\mathbf{y}^{\prime}}}\right)_{\mathbf{x}^{\prime},t}=\left(\mathbf{\tilde{v}}_{L}\overleftarrow{\nabla_{\mathbf{\mathbf{y}}}}\right)\cdot\tilde{\mathbf{F}}$.
Recall our convention to drop the subscript $\mathbf{y}$ in the tensor
$\nabla\mathbf{\tilde{v}}_{L}$.%
}
\begin{equation}
\dfrac{d^{L}\tilde{\mathbf{F}}}{dt}=\left(\mathbf{\tilde{v}}_{L}\overleftarrow{\nabla}\right)\cdot\tilde{\mathbf{F}}.\label{eq:19}
\end{equation}
Eq.(\ref{eq:17-3-1}) becomes
\begin{eqnarray}
\dfrac{d^{L}u}{dt} & = & T\dfrac{d^{L}s}{dt}+\sum_{i=1}^{n}M_{i}^{*}\dfrac{d^{L}n_{i}}{dt}+\sum_{k=1}^{K}\Phi_{k}^{*}\dfrac{d^{L}\varphi_{k}}{dt}+\nabla\cdot\left(\sum_{i=1}^{n}\dfrac{\partial u}{\partial\nabla n_{i}}\dfrac{d^{L}n_{i}}{dt}+\sum_{k=1}^{K}\dfrac{\partial u}{\partial\nabla\varphi_{k}}\dfrac{d^{L}\varphi_{k}}{dt}\right)\nonumber \\
 & - & \left(\sum_{i=1}^{n}\dfrac{\partial u}{\partial\nabla n_{i}}\nabla n_{i}+\sum_{k=1}^{K}\dfrac{\partial u}{\partial\nabla\varphi_{k}}\nabla\varphi_{k}\right):\nabla\mathbf{v}_{L}+\left(\tilde{\mathbf{\boldsymbol{\sigma}}}-\omega\mathbf{I}\right):\nabla\mathbf{\tilde{v}}_{L}.\label{eq:20}
\end{eqnarray}
Note that this equation contains both the total (marker network) velocity
and the local velocity of the lattice, the former coming from the
material time derivatives and the latter from the local lattice deformation
gradient. 

The term with the chemical potentials can be further rearranged using
the species conservation law, Eq.(\ref{eq:8}):
\begin{eqnarray}
\sum_{i=1}^{n}M_{i}^{*}\dfrac{dn_{i}}{dt} & = & -\sum_{i=1}^{n}M_{i}^{*}\nabla\cdot\mathbf{J}_{i}^{L}-\left(\sum_{i=1}^{n}M_{i}^{*}n_{i}\right)\nabla\cdot\mathbf{v}_{L}\nonumber \\
 & = & -\nabla\cdot\left(\sum_{i=1}^{n}M_{i}^{*}\mathbf{J}_{i}^{L}\right)+\sum_{i=1}^{n}\mathbf{J}_{i}^{L}\cdot\nabla M_{i}^{*}-\left(\sum_{i=1}^{n}M_{i}^{*}n_{i}\right)\nabla\cdot\mathbf{v}_{L}.\label{eq:22}
\end{eqnarray}
For further calculations we need the energy and entropy rates to appear
in the combinations $(d^{L}u/dt+u\nabla\cdot\mathbf{v}_{L})$ and
$(d^{L}s/dt+s\nabla\cdot\mathbf{v}_{L})$, respectively. This is readily
achieved by adding and subtracting $u\nabla\cdot\mathbf{v}_{L}$ and
$s\nabla\cdot\mathbf{v}_{L}$ in Eq.(\ref{eq:20}). After simple rearrangements
we arrive at the equation

\begin{eqnarray}
 & \dfrac{d^{L}u}{dt}+u\nabla\cdot\mathbf{v}_{L}= & T\left(\dfrac{d^{L}s}{dt}+s\nabla\cdot\mathbf{v}_{L}\right)+\sum_{i=1}^{n}\mathbf{J}_{i}^{L}\cdot\nabla M_{i}^{*}+\sum_{k=1}^{K}\Phi_{k}^{*}\dfrac{d^{L}\varphi_{k}}{dt}\nonumber \\
 &  & +\nabla\cdot\left(\sum_{i=1}^{n}\dfrac{\partial u}{\partial\nabla n_{i}}\dfrac{d^{L}n_{i}}{dt}+\sum_{k=1}^{K}\dfrac{\partial u}{\partial\nabla\varphi_{k}}\dfrac{d^{L}\varphi_{k}}{dt}-\sum_{i=1}^{n}M_{i}^{*}\mathbf{J}_{i}^{L}\right)\nonumber \\
 &  & +\left(\tilde{\mathbf{\boldsymbol{\sigma}}}+\mathbf{A}^{*}\right):\nabla\mathbf{v}_{L}-\left(\tilde{\mathbf{\boldsymbol{\sigma}}}-\omega\mathbf{I}\right):\mathbf{R},\label{eq:35}
\end{eqnarray}
where 
\begin{equation}
\mathbf{R}\equiv\left(\nabla\mathbf{v}_{L}-\nabla\mathbf{\tilde{v}}_{L}\right).\label{eq:24-1}
\end{equation}

The tensor $\mathbf{R}$ represents the permanent part of the total
deformation rate coming from the site generation and annihilation.
According to Eq.(\ref{eq:6-9}) its trace,
\begin{equation}
\textrm{Tr}(\mathbf{R})=\left(\nabla\cdot\mathbf{v}_{L}-\nabla\cdot\mathbf{\tilde{v}}_{L}\right)=\dfrac{r_{s}}{n_{s}},\label{eq:25}
\end{equation}
gives the site generation rate $r_{s}$. However, the tensor $\mathbf{R}$
carries more information than $r_{s}$ as it reflects the possible
anisotropy in the generation of lattice sites. It differentiates,
for example, between insertion of new lattice planes normal to a certain
direction and creation of the same number of sites by uniform ``swelling''
of the material. In fact, $\mathbf{R}$ captures even a pure shear
deformation rate in which new lattice planes are inserted parallel
to one crystallographic orientation and simultaneously removed parallel
to another crystallographic orientation perpendicular to the first,
so that the total number of sites remains constant. One possible mechanism
of this process would be a concurrent climb of two perpendicular sets
of edge dislocations, one inserting lattice planes and the other eliminating
perpendicular lattice planes. This could be accomplished by vacancy
diffusion between the cores of the two dislocation sets without changing
the net amount of vacancies in the region.

Tensor $\mathbf{R}$ is related to the generalized creep strain-rate
tensor $\boldsymbol{\varepsilon}_{gc}$ introduced by Svoboda \emph{et
al.}\cite{Svoboda2006,Fischer2011} although the latter, by contrast
to $\mathbf{R}$, comprises both permanent and elastic parts of the
deformation. Similar to $\mathbf{R}$, the tensor $\boldsymbol{\varepsilon}_{gc}$
includes both the volume creep deformation by ``swelling'' or contraction
and shear deformation arising due to orientational anisotropy of the
microstructure or from non-hydrostatic components of the stress tensor.

It should be emphasized that Eq.(\ref{eq:35}) has been derived from
the fundamental Eq.(\ref{eq:16}) by a chain of mathematical transformations
without any additional physical assumptions or approximations other
than the conservation and balance equations of Secs.~\ref{sec:Kinematics}
and \ref{sec:Balance-equations}. Equation (\ref{eq:35}) represents
a time-dependent form of the first and second laws of thermodynamics
for \emph{reversible} processes in a continuous medium with the postulated
equation of state (\ref{eq:16}).

\section{The state of equilibrium\label{sec:Equilibrium-state}}

\subsection{Derivation of equilibrium conditions\label{sub:Equilibrium}}

Before analyzing irreversible processes, we will derive the conditions
of thermodynamic equilibrium of a multicomponent solid capable of
site generation. This could be done by requiring that the first-order
variation of the total energy of a given material region enclosed
in a rigid envelope be zero under the constraints of fixed entropy
and fixed total number of particles of every species. Instead of considering
infinitesimal variations of the relevant parameters, we will reuse
Eq.(\ref{eq:35}) by treating the rates of the reversible changes
of those parameters as their variations. For example, the variation
$\delta\varphi_{k}$ can be formally considered to occur per unit
time and be represented by the material derivative $d^{L}\varphi_{k}/dt$.
Likewise, the virtual lattice displacement $\delta\mathbf{x}_{L}$
can be thought of as occurring per unit time and be replaced by the
lattice velocity $\mathbf{v}_{L}$. The macroscopic kinetic energy
is a second-order variation and is excluded. This treatment is completely
equivalent to the virtual displacement method usually applied for
finding thermodynamic equilibrium of continuous media.\cite{Malvern69,Larche73,Voorhees2004}

The equilibrium condition is
\begin{equation}
\int\left(\dfrac{d^{L}u}{dt}+u\nabla\cdot\mathbf{v}_{L}\right)dV+\int\left(\dfrac{d^{L}\psi}{dt}+\psi\nabla\cdot\mathbf{v}_{L}\right)dV-\lambda\int\left(\dfrac{d^{L}s}{dt}+s\nabla\cdot\mathbf{v}_{L}\right)dV=0.\label{eq:22-1}
\end{equation}
The first integral is equivalent to
\begin{equation}
\int\left(\dfrac{\partial u^{\prime}}{\partial t}\right)_{\mathbf{x}^{\prime}}dV^{\prime},\label{eq:23-2}
\end{equation}
$u^{\prime}$ being internal entropy per unit reference volume,%
\footnote{Indeed, using $u^{\prime}=Gu$ with $G\equiv\det\mathbf{F}$ we have
$(\partial u^{\prime}/\partial t)_{\mathbf{x}^{\prime}}=u(\partial G/\partial t)_{\mathbf{x}^{\prime}}+G(\partial u/\partial t)_{\mathbf{x}^{\prime}}$,
which with the help of $(\partial G/\partial t)_{\mathbf{x}^{\prime}}=G\nabla\cdot\mathbf{v}_{L}$
and $(\partial u/\partial t)_{\mathbf{x}^{\prime}}=(\partial u/\partial t)_{\mathbf{x}}+\mathbf{v}_{L}\cdot\nabla u$
becomes $(\partial u^{\prime}/\partial t)_{\mathbf{x}^{\prime}}=G\left(d^{L}u/dt+u\nabla\cdot\mathbf{v}_{L}\right)$.%
} and represents the rate of internal energy change of a given material
region defined by lattice markers. Likewise, the second and third
integrals represent the rates of potential energy and entropy changes
of the same material region. The entropy integral has been added to
impose the entropy constraint using the Lagrange multiplier $\lambda$.
The required conservation of the total amount of each species will
be enforced by zero normal components of the diffusion fluxes at the
boundary of the region and need not be imposed via additional Lagrange
multipliers.

Inserting the first integrand from Eq.(\ref{eq:35}), the divergence
term becomes the surface integral over the boundary of the region,
\begin{equation}
\int\mathbf{n}\cdot\left(\sum_{i=1}^{n}\dfrac{\partial u}{\partial\nabla n_{i}}\dfrac{d^{L}n_{i}}{dt}+\sum_{k=1}^{K}\dfrac{\partial u}{\partial\nabla\varphi_{k}}\dfrac{d^{L}\varphi_{k}}{dt}-\sum_{i=1}^{n}M_{i}^{*}\mathbf{J}_{i}^{L}\right)dA,\label{eq:24-2}
\end{equation}
$\mathbf{n}$ being a unit normal vector pointing outside the region
and $dA$ an increment of area. To ensure that the region is closed,
the normal components of the diffusion fluxes will be taken to be
zero, $\mathbf{n}\cdot\mathbf{J}_{i}^{L}=0$. Imposing also fixed
boundary values of $n_{i}$ and $\varphi_{k}$, this surface integral
vanishes. Furthermore, the volume integral
\begin{equation}
\int\left(\tilde{\mathbf{\boldsymbol{\sigma}}}+\mathbf{A}^{*}\right):\nabla\mathbf{v}_{L}dV\label{eq:25-1}
\end{equation}
can be rewritten using the divergence theorem as
\begin{equation}
\int\mathbf{n}\cdot\left(\tilde{\mathbf{\boldsymbol{\sigma}}}+\mathbf{A}^{*}\right)\cdot\mathbf{v}_{L}dA-\int\nabla\cdot\left(\tilde{\mathbf{\boldsymbol{\sigma}}}+\mathbf{A}^{*}\right)\cdot\mathbf{v}_{L}dV,\label{eq:26}
\end{equation}
where the surface integral is zero due to the boundary condition $\mathbf{v}_{L}=\mathbf{0}$
(rigid boundary). Similarly, per Eq.(\ref{eq:13-3}) the potential
energy integral contains the potential energy flux which vanishes
on the boundary, leaving 
\begin{equation}
\int\left(\dfrac{d^{L}\psi}{dt}+\psi\nabla\cdot\mathbf{v}_{L}\right)dV=-\int\left(\sum_{i=1}^{n}\mathbf{b}_{i}\cdot\mathbf{J}_{i}^{L}+\mathbf{b}\cdot\mathbf{v}_{L}\right)dV.\label{eq:26-1}
\end{equation}
 Combining the above equations, Eq.(\ref{eq:22-1}) becomes
\begin{eqnarray}
 &  & \int\left(\dfrac{d^{L}s}{dt}+s\nabla\cdot\mathbf{v}_{L}\right)(T-\lambda)dV+\int\sum_{i=1}^{n}\mathbf{J}_{i}^{L}\cdot\nabla\left(M_{i}^{*}+\psi_{i}\right)dV+\int\sum_{k=1}^{K}\Phi_{k}^{*}\dfrac{d^{L}\varphi_{k}}{dt}dV\nonumber \\
 &  & -\int\left[\nabla\cdot\left(\tilde{\mathbf{\boldsymbol{\sigma}}}+\mathbf{A}^{*}\right)+\mathbf{b}\right]\cdot\mathbf{v}_{L}dV-\int\left(\tilde{\mathbf{\boldsymbol{\sigma}}}-\omega\mathbf{I}\right):\mathbf{R}dV=0.\label{eq:35-2}
\end{eqnarray}

In the state of equilibrium this relation must hold for any arbitrarily
chosen region of the material. The integrands are proportional to
the entropy rate $(d^{L}s/dt+s\nabla\cdot\mathbf{v}_{L})$, the phase-field
rates $d^{L}\varphi_{k}/dt$, the diffusion fluxes $\mathbf{J}_{i}^{L}$,
the lattice velocity $\mathbf{v}_{L}$, and the creep deformation
rate $\mathbf{R}$, respectively. All these rates represent independent
variations away from the equilibrium state. Assuming that they can
take any arbitrary positive or negative values, the coefficients multiplying
these rates must be zero. We thus arrive at the following set of necessary
conditions of equilibrium:
\begin{equation}
T=\lambda=\textnormal{const}\quad\quad\textnormal{Thermal equilibrium}\label{eq:71-1}
\end{equation}
\begin{equation}
M_{i}^{*}+\psi_{i}=\textnormal{const}\quad\quad\textnormal{Chemical equilibrium }\label{eq:73}
\end{equation}
\begin{equation}
\Phi_{k}^{*}=0\quad\quad\textnormal{Phase-field equilibrium }\label{eq:74}
\end{equation}
\begin{equation}
\nabla\cdot\left(\tilde{\mathbf{\boldsymbol{\sigma}}}+\mathbf{A}^{*}\right)+\mathbf{b}=0\quad\quad\textnormal{Mechanical equilibrium }\label{eq:75}
\end{equation}
\begin{equation}
\tilde{\mathbf{\boldsymbol{\sigma}}}=\omega\mathbf{I}.\quad\quad\textnormal{Site generation equilibrium }\label{eq:76}
\end{equation}

\subsection{Discussion of the equilibrium conditions\label{sub:Equilibrium-2}}

Eqs.(\ref{eq:71-1})-(\ref{eq:74}) reproduce the well-known conditions
of thermal, chemical and phase field equilibrium: the uniformity of
the temperature field, the constancy of the non-classical chemical
potential $M_{i}^{*}$ plus the external potential $\psi_{i}$ for
every species, and vanishing variational derivative $\Phi_{k}^{*}$
for every phase field. 

The mechanical equilibrium condition could have been obtained from
zero accelerations and zero diffusion fluxes in the momentum balance
equation (\ref{eq:12-2}), giving $\nabla\cdot\boldsymbol{\sigma}+\mathbf{b}=\mathbf{0}$.
Equation (\ref{eq:75}) shows that tensor $(\tilde{\mathbf{\boldsymbol{\sigma}}}+\mathbf{A}^{*})$
plays the role of the non-classical stress tensor. The latter has
long been known in fluid systems as the capillary tensor or Korteweg
stress.\cite{Korteweg1901} In classical regions where the gradients
of the chemical composition and phase fields can be neglected and
thus $\mathbf{\mathbf{A}^{*}}=\mathbf{0}$, the mechanical equilibrium
condition reduces to $\nabla\cdot\tilde{\mathbf{\boldsymbol{\sigma}}}+\mathbf{b}=\mathbf{0}$,
confirming that the tensor $\tilde{\mathbf{\boldsymbol{\sigma}}}$
defined earlier by Eq.(\ref{eq:16-2}) is indeed the equilibrium stress
tensor in a classical solid.

Equation (\ref{eq:76}) is the condition of equilibrium with respect
to site generation and annihilation, stating that tensor $\tilde{\mathbf{\boldsymbol{\sigma}}}$
must be diagonal: $\tilde{\mathbf{\boldsymbol{\sigma}}}\equiv-p\mathbf{I}$.
This condition must be fulfilled everywhere in the equilibrium material,
including non-classical regions with significant gradients of $n_{i}$
and/or $\varphi_{k}$, such as interface regions. However, the actual
stress tensor in such regions, $(-p\mathbf{I}+\mathbf{A}^{*})$, remains
non-hydrostatic due to the non-classical contribution $\mathbf{A}^{*}$. 

If Eq.(\ref{eq:76}) is satisfied, the mechanical equilibrium condition
becomes
\begin{equation}
-\nabla p+\nabla\cdot\mathbf{A}^{*}+\mathbf{b}=\mathbf{0}\label{eq:76-1}
\end{equation}
and in classical regions reduces to the standard hydrostatic equilibrium
condition $\nabla p=\mathbf{b}$.\cite{Malvern69} Thus, in the presence
of efficient sinks and sources of vacancies capable of maintaining
site equilibrium the solid behaves rheologically like a fluid.

Note that by inserting the obtained equilibrium conditions (\ref{eq:71-1})-(\ref{eq:74})
and (\ref{eq:76}) in the generalized Gibbs-Duhem equation (\ref{eq:17-6})
we immediately recover Eq.(\ref{eq:18-1}) or its equivalent form
(\ref{eq:76-1}). In other words, if all other equilibrium conditions
are satisfied, the mechanical equilibrium condition follows from the
generalized Gibbs-Duhem equation (\ref{eq:17-6}). The reverse is
not true: the mechanical equilibrium condition (\ref{eq:75}) can
be satisfied even if the material has not reached complete equilibrium,
in which case Eq.(\ref{eq:18-1}) is invalid.

According to Eq.(\ref{eq:76}), in equilibrium $p=-\omega$, i.e.,
\begin{equation}
u-Ts+p-\sum_{i=1}^{n}M_{i}n_{i}=0,\label{eq:77}
\end{equation}
in both classical and non-classical regions. In classical regions
this relation has a clear thermodynamic meaning. In such regions the
actual state of stress of the material is hydrostatic with the pressure
$p$. Under such conditions one can uniquely define the chemical potentials
$\mu_{i}$ of all chemical species as well as the chemical potential
$\mu_{v}$ of vacancies treated as fictitious massless species.%
\footnote{In non-hydrostatically stressed solids chemical potentials of material
species and vacancies cannot be defined simultaneously due to the
network constraint.\cite{Larche73,Larche_Cahn_78,Larche1985}%
} The diffusion potentials $M_{i}$ are then $M_{i}=\mu_{i}-\mu_{v}$
and the left-hand side of Eq.(\ref{eq:77}) becomes
\begin{equation}
0=\left(u-Ts+p-\sum_{i=1}^{n}\mu_{i}n_{i}-\mu_{v}n_{v}\right)+\mu_{v}n_{s}=\mu_{v}n_{s},\label{eq:78}
\end{equation}
where $n_{v}\equiv n_{s}-\sum_{i}n_{i}$ is the number density of
vacancies per unit volume and we used the Gibbs relation for hydrostatic
systems, \cite{Willard_Gibbs}
\begin{equation}
u-Ts+p=\sum_{i=1}^{n}\mu_{i}n_{i}+\mu_{v}n_{v}.\label{eq:79}
\end{equation}
Thus, Eq.(\ref{eq:77}) predicts that the equilibrium chemical potential
of vacancies in classical regions is zero:
\begin{equation}
\mu_{v}=0.\label{eq:80}
\end{equation}
This relation cannot be extended to non-classical regions, e.g. interfaces,
where $\mu_{v}$ remains undefined.

It is important to recognize that the equilibrium condition (\ref{eq:76})
has been derived by considering \emph{independent} variations of \emph{all}
components of the creep deformation rate tensor $\mathbf{R}$. Under
real conditions the material's microstructure can impose restrictions
on some of such variations. For example, the material can be only
capable of site generation/annihilation by insertion or removal of
lattice planes normal to a particular direction, e.g. by growth or
shrinkage of extrinsic stacking faults in those planes. Alternatively,
the site generation/annihilation can occur exclusively by growth or
dissolution of nano-pores permitting only isotropic ``swelling'' or
contraction of the material. In all such cases the material can reach
a constrained thermodynamic equilibrium with only some of the components
of $(\tilde{\mathbf{\boldsymbol{\sigma}}}-\omega\mathbf{I})$, or
their linear combinations, being zero. In such cases the equilibrium
stress tensor $\tilde{\mathbf{\boldsymbol{\sigma}}}$ need not be
hydrostatic. Under such constrained equilibrium conditions Eq.(\ref{eq:80})
is no longer valid, and furthermore, $\mu_{v}$ itself is undefined. 

In the limiting case when the material does not contain any sinks
or sources of vacancies, $\mathbf{R}$ is identically zero and the
material can be equilibrated in any non-hydrostatic state of stress.
Equation (\ref{eq:76}) should be then removed from the list of equilibrium
conditions.

\section{Irreversible thermodynamics\label{sec:Irreversible-TD}}

\subsection{The entropy production rate}

As indicated in Sec.~\ref{sub:Entropy-balance}, a route to the entropy
production is to (i) insert in Eq.(\ref{eq:35}) the internal energy
rate $(d^{L}u/dt+u\nabla\cdot\mathbf{v}_{L})$ from the energy balance
equation (\ref{eq:13}), and (ii) split the obtained total entropy
rate $(d^{L}s/dt+s\nabla\cdot\mathbf{v}_{L})$ into the entropy flux
$-\nabla\cdot\mathbf{J}_{s}^{L}$ and entropy production rate $\dot{s}$. 

Step (i) gives
\begin{eqnarray}
 &  & T\left(\dfrac{d^{L}s}{dt}+s\nabla\cdot\mathbf{v}_{L}\right)+\sum_{i=1}^{n}\mathbf{J}_{i}^{L}\cdot\left\{ \nabla\left[M_{i}^{*}+\dfrac{m_{i}}{2n_{i}^{2}}\left(\mathbf{J}_{i}\cdot\mathbf{J}_{i}\right)\right]-\mathbf{b}_{i}+m_{i}\dfrac{d^{L}\mathbf{v}_{i}}{dt}\right\} \nonumber \\
 &  & +\nabla\cdot\left(\mathbf{J}_{u}^{L}+\sum_{i=1}^{n}\dfrac{\partial u}{\partial\nabla n_{i}}\dfrac{d^{L}n_{i}}{dt}+\sum_{k=1}^{K}\dfrac{\partial u}{\partial\nabla\varphi_{k}}\dfrac{d^{L}\varphi_{k}}{dt}-\sum_{i=1}^{n}M_{i}^{*}\mathbf{J}_{i}^{L}\right)\nonumber \\
 &  & +\sum_{k=1}^{K}\Phi_{k}^{*}\dfrac{d^{L}\varphi_{k}}{dt}-\left(\boldsymbol{\sigma}-\mathbf{M}-\tilde{\mathbf{\boldsymbol{\sigma}}}-\mathbf{A}^{*}\right):\nabla\mathbf{v}_{L}-\left(\tilde{\mathbf{\boldsymbol{\sigma}}}-\omega\mathbf{I}\right):\mathbf{R}=0.\label{eq:35-1}
\end{eqnarray}
Solving this equation for the total entropy rate,
\begin{eqnarray}
\dfrac{d^{L}s}{dt}+s\nabla\cdot\mathbf{v}_{L} & = & -\nabla\cdot\left(\dfrac{\mathbf{J}_{q}^{L}}{T}\right)-\dfrac{1}{T}\sum_{i=1}^{n}\mathbf{J}_{i}^{L}\cdot\left\{ \nabla\left[M_{i}^{*}+\dfrac{m_{i}}{2n_{i}^{2}}\left(\mathbf{J}_{i}\cdot\mathbf{J}_{i}\right)\right]-\mathbf{b}_{i}+m_{i}\dfrac{d^{L}\mathbf{v}_{i}}{dt}\right\} \nonumber \\
 &  & -\dfrac{1}{T}\sum_{k=1}^{K}\Phi_{k}^{*}\dfrac{d^{L}\varphi_{k}}{dt}+\mathbf{J}_{q}^{L}\cdot\nabla\dfrac{1}{T}\nonumber \\
 &  & +\dfrac{1}{T}\left(\boldsymbol{\sigma}-\mathbf{M}-\tilde{\mathbf{\boldsymbol{\sigma}}}-\mathbf{A}^{*}\right):\nabla\mathbf{v}_{L}+\dfrac{1}{T}\left(\tilde{\mathbf{\boldsymbol{\sigma}}}-\omega\mathbf{I}\right):\mathbf{R},\label{eq:52}
\end{eqnarray}
where
\begin{eqnarray}
\mathbf{J}_{q}^{L} & \equiv & \mathbf{J}_{u}^{L}+\sum_{i=1}^{n}\dfrac{\partial u}{\partial\nabla n_{i}}\dfrac{d^{L}n_{i}}{dt}+\sum_{k=1}^{K}\dfrac{\partial u}{\partial\nabla\varphi_{k}}\dfrac{d^{L}\varphi_{k}}{dt}-\sum_{i=1}^{n}M_{i}^{*}\mathbf{J}_{i}^{L}\label{eq:53}
\end{eqnarray}
is a heat flux relative to the lattice. The latter equals the total
internal energy flux $\mathbf{J}_{u}^{L}$ less the internal energy
transferred by diffusion and by the motion of phase transformation
fronts or GBs.

Identifying the entropy flux
\begin{equation}
\mathbf{J}_{s}^{L}\equiv\dfrac{\mathbf{J}_{q}^{L}}{T}\label{eq:53-1}
\end{equation}
we finally obtain the entropy production rate
\begin{eqnarray}
\dot{s} & = & -\dfrac{1}{T^{2}}\nabla T\cdot\mathbf{J}_{q}^{L}-\dfrac{1}{T}\sum_{i=1}^{n}\left\{ \nabla\left[M_{i}^{*}+\dfrac{m_{i}}{2n_{i}^{2}}\left(\mathbf{J}_{i}\cdot\mathbf{J}_{i}\right)+\psi_{i}\right]+m_{i}\dfrac{d^{L}\mathbf{v}_{i}}{dt}\right\} \cdot\mathbf{J}_{i}^{L}\nonumber \\
 &  & -\dfrac{1}{T}\sum_{k=1}^{K}\Phi_{k}^{*}\dfrac{d^{L}\varphi_{k}}{dt}+\dfrac{1}{T}\left(\boldsymbol{\sigma}-\mathbf{M}-\tilde{\mathbf{\boldsymbol{\sigma}}}-\mathbf{A}^{*}\right):\nabla\mathbf{v}_{L}+\dfrac{1}{T}\left(\tilde{\mathbf{\boldsymbol{\sigma}}}-\omega\mathbf{I}\right):\mathbf{R}.\label{eq:84-1}
\end{eqnarray}

The individual terms of Eq.(\ref{eq:84-1}) describe the entropy production
due to: (i) heat conduction; (ii) diffusion driven by gradients of
the non-classical diffusion potentials $M_{i}^{*}$, kinetic energy
of diffusion $m_{i}(\mathbf{J}_{i}\cdot\mathbf{J}_{i})/2n_{i}^{2}$
and external potentials $\psi_{i}$, and by inertia forces $m_{i}d^{L}\mathbf{v}_{i}/dt$;
(iii) evolution of the phase fields, (iv) viscous dissipation by conversion
of the strain rate to heat (e.g., generation of phonons), and (v)
generation/annihilation of lattice sites. Each term can be interpreted
as the product of a driving force and a conjugate generalized ``flux'',
the ``fluxes'' being $\mathbf{J}_{q}^{L}$ (heat), $\mathbf{J}_{i}^{L}$
(diffusion), $d^{L}\varphi_{k}/dt$ (phase-field evolution rate),
$\nabla\mathbf{v}_{L}$ (deformation rate) and $\mathbf{R}$ (site
generation rate).

Equation (\ref{eq:84-1}) represents the exact entropy production.
For applications to slow processes such as creep, it can be simplified
by neglecting the terms quadratic in diffusion fluxes and the inertia
terms (see Appendix B). The approximate form of the entropy production
rate, which will be used in the rest of the paper, becomes 

\begin{eqnarray}
\dot{s} & = & -\dfrac{1}{T^{2}}\nabla T\cdot\mathbf{J}_{q}^{L}-\dfrac{1}{T}\sum_{i=1}^{n}\nabla\left(M_{i}^{*}+\psi_{i}\right)\cdot\mathbf{J}_{i}^{L}\nonumber \\
 &  & -\dfrac{1}{T}\sum_{k=1}^{K}\Phi_{k}^{*}\dfrac{d^{L}\varphi_{k}}{dt}+\dfrac{1}{T}\left(\boldsymbol{\sigma}-\tilde{\mathbf{\boldsymbol{\sigma}}}-\mathbf{A}^{*}\right):\nabla\mathbf{v}_{L}+\dfrac{1}{T}\left(\tilde{\mathbf{\boldsymbol{\sigma}}}-\omega\mathbf{I}\right):\mathbf{R}.\label{eq:84}
\end{eqnarray}

It is instructive to apply Eq.(\ref{eq:84}) to the state of thermodynamic
equilibrium, in which all driving forces must vanish. Equating the
driving forces to zero recovers the previously found conditions of
thermal equilibrium (\ref{eq:71-1}), chemical equilibrium (\ref{eq:73}),
phase field equilibrium (\ref{eq:74}), and the site generation equilibrium
(\ref{eq:76}) (Sec.~\ref{sub:Equilibrium}). Thus, the fully equilibrated
material is correctly predicted to be hydrostatic. According to Eq.(\ref{eq:84}),
in the absence of viscous dissipation the dynamic stress tensor $\boldsymbol{\sigma}$
reduces to its static value $(\tilde{\mathbf{\boldsymbol{\sigma}}}+\mathbf{A}^{*})$
(Sec.~\ref{sub:Equilibrium-2}):
\begin{equation}
\boldsymbol{\sigma}=\tilde{\mathbf{\boldsymbol{\sigma}}}+\mathbf{A}^{*}.\quad\quad\textnormal{No viscous dissipation }\label{eq:83}
\end{equation}
We do not recover the mechanical equilibrium condition (\ref{eq:75}).
However, the latter follows at once from the generalized Gibbs-Duhem
equation (\ref{eq:17-6}) (see Sec.~\ref{sub:Gibbs-Duhem}) provided
that all other equilibrium conditions are satisfied. 

It is interesting to note that if any of the components of $\left(\tilde{\mathbf{\boldsymbol{\sigma}}}-\omega\mathbf{I}\right)$
in the equilibrium state are nonzero due to restrictions on site generation,
the condition of zero entropy production does not recover the mechanical
equilibrium condition. This should not be surprising since $\dot{s}=0$
is only a necessary but not sufficient condition of thermodynamic
equilibrium. The absence of entropy production can be satisfied not
only in the equilibrium state but also during (nearly) reversible
mechanical processes, such as propagation of elastic waves with negligible
dissipation.

In many situations some of the driving forces appearing in Eq.(\ref{eq:84})
can be negligibly small and the process can be driven by the remaining
forces. For example, on sufficiently short time scales the site generation
and diffusion processes can be neglected ($\mathbf{R}=\mathbf{0}$,
$\mathbf{J}_{i}^{L}=\mathbf{0}$) and the material can undergo fast
(e.g., shock) deformation accompanied by viscous dissipation, conduction
of heat and possibly diffusionless phase transformations. As another
example, for slow enough processes one can neglect the viscous dissipation
and assume thermal and mechanical equilibrium, leaving only diffusion,
phase transformations or GB motion, and site generation as the dominant
processes. It is this latter regime that appears to be most relevant
to diffusional creep and will be discussed in more detail later in
Section \ref{sec:Simple-model}.

\subsection{Phenomenological relations}

\subsubsection{Material symmetry considerations}

We will next postulate linear phenomenological relations between the
fluxes and forces appearing in the entropy production rate, Eq.(\ref{eq:84}).
Generally, each flux can be linearly related (coupled) to all forces
entering this expression, and the matrix of the linear coefficients
must be symmetric by the Onsager reciprocal relations.\cite{Onsager1931a,Onsager1931b}
It is known, however, that symmetry properties of the material can
prevent coupling between certain fluxes and forces (Curie symmetry
principle). In particular, if all properties of the material are isotropic,
a flux can be caused only by forces having the same tensorial character.
Quantities with four distinct types of tensorial character usually
occur in expressions for the entropy production: scalars, polar vectors,
axial vectors, and symmetric traceless tensors of rank two. We will
start by rearranging the terms in (\ref{eq:84}) according to their
tensorial character. This step requires only mathematical rearrangements
in Eq.(\ref{eq:84}) and does not involve any assumptions regarding
the symmetry of the material.

The phase field rates $d^{L}\varphi_{k}/dt$ are scalars and the fluxes
of the chemical components and heat are polar vectors. The forces
conjugate to these fluxes have the same tensorial character as the
fluxes. Thus we need not do anything about these terms. The remaining
terms are double-contractions of second rank tensors, which will be
partitioned as follows.\cite{De-Groot1984} Each tensor $\mathbf{A}$
is split in three parts,
\begin{equation}
\mathbf{A}=\dfrac{1}{3}\textrm{Tr}(\mathbf{A})\mathbf{I}+\mathbf{A}^{(s)}+\mathbf{A}^{(a)},\label{eq:85}
\end{equation}
where
\begin{equation}
\mathbf{A}^{(s)}=\dfrac{1}{2}\left(\mathbf{A}+\mathbf{A}^{T}\right)-\dfrac{1}{3}\textrm{Tr}(\mathbf{A})\mathbf{I}\label{eq:86}
\end{equation}
is the traceless symmetric part of $\mathbf{A}$ and 
\begin{equation}
\mathbf{A}^{(a)}=\dfrac{1}{2}\left(\mathbf{A}-\mathbf{A}^{T}\right)\label{eq:87}
\end{equation}
is the anti-symmetric part of $\mathbf{A}$. Applying this decomposition
to two second-rank tensors $\mathbf{A}$ and $\mathbf{B}$, it can
be shown that
\begin{equation}
\mathbf{A}:\mathbf{B}=\dfrac{1}{3}\textrm{Tr}(\mathbf{A})\textrm{Tr}(\mathbf{B})+\mathbf{A}^{(s)}:\mathbf{B}^{(s)}+\mathbf{A}^{(a)}:\mathbf{B}^{(a)}.\label{eq:88}
\end{equation}
The last term is equivalent to a dot product of two axial vectors.\cite{De-Groot1984}
Thus, the operation ``$:$'' only couples parts of the tensors that
have the same tensorial character. 

Applying this tensor decomposition and grouping together the terms
with the same tensor character, the entropy production rate becomes

\begin{eqnarray}
\dot{s} & = & -\dfrac{1}{T}\sum_{k=1}^{K}\Phi_{k}^{*}\dfrac{d^{L}\varphi_{k}}{dt}+\dfrac{1}{T}\Pi\nabla\cdot\mathbf{v}_{L}-\dfrac{1}{T}\left(\omega-\tilde{\sigma}_{h}\right)r\nonumber \\
 & - & \dfrac{1}{T^{2}}\nabla T\cdot\mathbf{J}_{q}^{L}-\dfrac{1}{T}\sum_{i=1}^{n}\nabla\left(M_{i}^{*}+\psi_{i}\right)\cdot\mathbf{J}_{i}^{L}\nonumber \\
 & + & \dfrac{1}{T}\boldsymbol{\tau}:\left(\nabla\mathbf{v}_{L}\right)^{(s)}+\dfrac{1}{T}\left(\tilde{\mathbf{\boldsymbol{\sigma}}}-\tilde{\sigma}_{h}\mathbf{I}\right):\mathbf{R}^{(s)}\nonumber \\
 & - & \dfrac{1}{T}\mathbf{A}^{(a)*}:\mathbf{W}.\label{eq:89}
\end{eqnarray}
where $r\equiv r_{s}/n_{s}$ is the number of new sites generated
per unit time per existing site and $\tilde{\sigma}_{h}=(1/3)\textrm{Tr}(\tilde{\mathbf{\boldsymbol{\sigma}}})$
is the ``hydrostatic part'' of $\tilde{\mathbf{\boldsymbol{\sigma}}}$.
In the above equation,
\begin{equation}
\left(\nabla\mathbf{v}_{L}\right)^{(s)}=\mathbf{D}-\dfrac{1}{3}\left(\nabla\cdot\mathbf{v}_{L}\right)\mathbf{I}\label{eq:90}
\end{equation}
is the total shear strain rate and
\begin{equation}
\mathbf{D}=\dfrac{1}{2}\left[\nabla\mathbf{v}_{L}+\left(\nabla\mathbf{v}_{L}\right)^{T}\right]\label{eq:91}
\end{equation}
is the deformation rate tensor.\cite{Malvern69} Tensor
\begin{equation}
\mathbf{W}\equiv\left(\nabla\mathbf{v}_{L}\right)^{(a)}=\dfrac{1}{2}\left[\nabla\mathbf{v}_{L}-\left(\nabla\mathbf{v}_{L}\right)^{T}\right]\label{eq:92}
\end{equation}
is sometimes called the vorticity tensor and characterizes the rate
of lattice rotation.\cite{Malvern69} The symmetric part of the creep
deformation rate $\mathbf{R}$ is
\begin{equation}
\mathbf{R}^{(s)}=\mathbf{D}-\tilde{\mathbf{D}}-\dfrac{r}{3}\mathbf{I},\label{eq:93}
\end{equation}
where
\begin{equation}
\tilde{\mathbf{D}}=\dfrac{1}{2}\left[\nabla\tilde{\mathbf{v}}_{L}+\left(\nabla\tilde{\mathbf{v}}_{L}\right)^{T}\right],\label{eq:93-1}
\end{equation}
and describes the rate of pure shear deformation caused by the creep
process. 

The scalar forces appearing in Eq.(\ref{eq:89}) include the non-classical
bulk viscosity stress
\begin{equation}
\Pi=\sigma_{h}-\tilde{\sigma}_{h}-\dfrac{1}{3}\textrm{Tr}\left(\mathbf{A^{*}}\right)\label{eq:94}
\end{equation}
and the volume driving force for the site generation, $(\omega-\tilde{\sigma}_{h})$.
The tensor forces include the non-classical viscous shear stress 
\begin{equation}
\boldsymbol{\tau}=\boldsymbol{\sigma}-\tilde{\mathbf{\boldsymbol{\sigma}}}-\mathbf{A}^{(s)*}-\left(\sigma_{h}-\tilde{\sigma}_{h}\right)\mathbf{I}\label{eq:95}
\end{equation}
and the driving force for the shear creep, $(\tilde{\mathbf{\boldsymbol{\sigma}}}-\tilde{\sigma}_{h}\mathbf{I})$.
The individual components of tensor $\mathbf{A}^{*}$ are 

\begin{equation}
\mathbf{A}^{(s)*}=-\sum_{i=1}^{n}\left(\dfrac{\partial u}{\partial\nabla n_{i}}\nabla n_{i}\right)^{(s)}-\sum_{k=1}^{K}\left(\dfrac{\partial u}{\partial\nabla\varphi_{k}}\nabla\varphi_{k}\right)^{(s)},\label{eq:96}
\end{equation}
\begin{equation}
\mathbf{A}^{(a)*}=-\sum_{i=1}^{n}\left(\dfrac{\partial u}{\partial\nabla n_{i}}\nabla n_{i}\right)^{(a)}-\sum_{k=1}^{K}\left(\dfrac{\partial u}{\partial\nabla\varphi_{k}}\nabla\varphi_{k}\right)^{(a)},\label{eq:97}
\end{equation}
\begin{equation}
\textrm{Tr}(\mathbf{A^{*}})\equiv3\left(\sum_{i=1}^{n}n_{i}\nabla\cdot\dfrac{\partial u}{\partial\nabla n_{i}}\right)-\sum_{i=1}^{n}\dfrac{\partial u}{\partial\nabla n_{i}}\cdot\nabla n_{i}-\sum_{k=1}^{K}\dfrac{\partial u}{\partial\nabla\varphi_{k}}\cdot\nabla\varphi_{k}.\label{eq:98}
\end{equation}

Note that the entropy production due to viscous dissipation is now
split in three parts: the bulk viscosity $\Pi\nabla\cdot\mathbf{v}_{L}$,
the viscous shear $\boldsymbol{\tau}:(\nabla\mathbf{v}_{L})^{(s)}$,
and the rotational viscosity $\mathbf{A}^{(a)*}:\mathbf{W}$. A similar
splitting is used for fluid systems.\cite{De-Groot1984} The site
generation is split in two parts: the volume part $(\omega-\tilde{\sigma}_{h})r$
describing isotropic ``swelling'' or shrinkage of the material, and
the shear part $(\tilde{\mathbf{\boldsymbol{\sigma}}}-\tilde{\sigma}_{h}\mathbf{I}):\mathbf{R}^{(s)}$
describing shape deformation without changing the total number of
sites. The latter process was discussed in the end of Sec.~\ref{sub:T-dependent-form}.

\subsubsection{Phenomenological relations for creep in isotropic materials}

Each of the four lines in Eq.(\ref{eq:89}) contains terms with contraction
of tensors of the same tensor character. If the material is isotropic,
only terms appearing in the same line can couple with each other but
not with terms in other lines.\cite{De-Groot1984} Furthermore, the
phenomenological coefficients have to be scalars regardless of the
tensor character of the fluxes and forces.%
\footnote{It is worth noting that this is true only when the symmetric tensors
are traceless. Symmetric tensors with a trace, such as the stress
and small-strain tensors in elastically isotropic (e.g, cubic) materials,
are linearly related with two phenomenological coefficients, e.g.,
the shear and bulk moduli.\cite{Nye-book}%
} This leads to the following phenomenological equations.

The scalar quantities appearing in the first line of Eq.(\ref{eq:89})
are coupled by the equations

\begin{eqnarray}
\dfrac{d^{L}\varphi_{k}}{dt} & = & -\dfrac{1}{T}\sum_{m=1}^{K}B_{km}\Phi_{m}^{*}+\dfrac{1}{T}B_{kv}\Pi-\dfrac{1}{T}B_{kr}\left(\omega-\tilde{\sigma}_{h}\right),\enskip\enskip k=1,...,K\nonumber \\
\nabla\cdot\mathbf{v}_{L} & = & \dfrac{1}{T}\sum_{m=1}^{K}B_{vm}\Phi_{m}^{*}+\dfrac{1}{T}B_{vv}\Pi-\dfrac{1}{T}B_{vr}\left(\omega-\tilde{\sigma}_{h}\right)\nonumber \\
r & = & -\dfrac{1}{T}\sum_{m=1}^{K}B_{rm}\Phi_{m}^{*}+\dfrac{1}{T}B_{rv}\Pi-\dfrac{1}{T}B_{rr}\left(\omega-\tilde{\sigma}_{h}\right).\label{eq:99}
\end{eqnarray}
By the Onsager relations,\cite{Onsager1931a,Onsager1931b} the $(K+2)\times(K+2)$
matrix $\mathbf{B}$ is symmetric and must be positive definite. In
particular, the diagonal coefficients $B_{kk}$, $B_{vv}$ and $B_{rr}$
must be positive. Generally, the site generation can be influenced
by viscous dissipation, phase transformations and GB motion. Conversely,
the phase field evolution and viscosity can be influenced by site
generation.

The second line of Eq.(\ref{eq:89}) describes diffusion of the chemical
species and heat. For simplicity, let us neglect the thermo-diffusion
cross-effects and decouple heat conduction from diffusion,
\begin{equation}
\mathbf{J}_{q}^{L}=-L_{qq}\dfrac{1}{T^{2}}\nabla T,\label{eq:100}
\end{equation}
where $L_{qq}>0$ is related to the heat conductivity coefficient
$\kappa$ by $\kappa=L_{qq}/T^{2}$. Then the diffusion equations
form a separate system,
\begin{equation}
\mathbf{J}_{i}^{L}=-\dfrac{1}{T}\sum_{j=1}^{n}L_{ij}\nabla\left(M_{j}^{*}+\psi_{j}\right),\enskip\enskip i=1,...,n.\label{eq:101}
\end{equation}
The $n\times n$ matrix $\mathbf{L}$ is symmetric and positive definite. 

In fluid systems in mechanical equilibrium, the chemical potential
gradients are linearly related to each other by the Gibbs-Duhem equation.\cite{Philibert,De-Groot1984}
As a result, one of the gradients can be eliminated. The complex solid
systems considered here follow the generalized Gibbs-Duhem equation
given by Eq.(\ref{eq:17-6}). Even in the absence of external fields
($\psi_{j}=0$), $\nabla M_{i}^{*}$ are linearly related only if
the materials is in thermal, phase-field, mechanical and site-generation
equilibrium (and thus in the hydrostatic state of stress). To keep
the treatment general, we will treat the diffusion potential gradients
as independent forces and the system of equations (\ref{eq:101})
as $n\times n$. 

From the third line of Eq.(\ref{eq:89}), the shear viscosity rate
and the shear creep deformation rate are coupled by the equations
\begin{eqnarray}
\mathbf{R}^{(s)} & = & \dfrac{1}{T}\mathcal{S}_{rr}\left(\tilde{\mathbf{\boldsymbol{\sigma}}}-\tilde{\sigma}_{h}\mathbf{I}\right)+\dfrac{1}{T}\mathcal{S}_{rv}\boldsymbol{\tau}\nonumber \\
\left(\nabla\mathbf{v}_{L}\right)^{(s)} & = & \dfrac{1}{T}\mathcal{S}_{vr}\left(\tilde{\mathbf{\boldsymbol{\sigma}}}-\tilde{\sigma}_{h}\mathbf{I}\right)+\dfrac{1}{T}\mathcal{S}_{vv}\boldsymbol{\tau},\label{eq:102}
\end{eqnarray}
where the matrix of coefficients is symmetric and positive definite.
The diagonal coefficients $\mathcal{S}_{rr}>0$ and $\mathcal{S}_{vv}>0$
characterize the kinetics of shear creep deformation and shear viscosity,
respectively, the latter being related to the viscosity coefficient
$\eta$ by $\mathcal{S}_{vv}=2T\eta$. Finally, from the fourth line
of Eq.(\ref{eq:89}) the rotational viscosity is decoupled from all
other effects and is described by the phenomenological relation
\begin{equation}
\left(\nabla\mathbf{v}_{L}\right)^{(a)}=-\mathcal{S}_{rot}\dfrac{1}{T}\mathbf{A}^{(a)*},\label{eq:103}
\end{equation}
where $\mathcal{S}_{rot}$ is related to the rotation viscosity coefficient
$\eta_{r}$ by $\eta_{r}=\mathcal{S}_{rot}/T$.

As already mentioned, for slow processes such as creep it is reasonable
to neglect the viscous dissipation and assume a uniform temperature
field and mechanical equilibrium. The remaining phenomenological equations
describe diffusion, phase-field evolution, site generation and creep
deformation. Assuming for simplicity that the material is not subject
to external fields, the obtained system of equations is 
\begin{equation}
\mathbf{J}_{i}^{L}=-\dfrac{1}{T}\sum_{j=1}^{n}L_{ij}\nabla M_{j}^{*},\enskip\enskip i=1,...,n\label{eq:104}
\end{equation}
\begin{eqnarray}
\dfrac{d^{L}\varphi_{k}}{dt} & = & -\dfrac{1}{T}\sum_{m=1}^{K}B_{km}\Phi_{m}^{*}-\dfrac{1}{T}B_{kr}\left(\omega-\tilde{\sigma}_{h}\right),\enskip\enskip k=1,...,K\label{eq:105}\\
r & = & -\dfrac{1}{T}\sum_{m=1}^{K}B_{rm}\Phi_{m}^{*}-\dfrac{1}{T}B_{rr}\left(\omega-\tilde{\sigma}_{h}\right),\label{eq:106}
\end{eqnarray}
\begin{equation}
\mathbf{R}^{(s)}=\dfrac{1}{T}\mathcal{S}_{rr}\left(\tilde{\mathbf{\boldsymbol{\sigma}}}-\tilde{\sigma}_{h}\mathbf{I}\right).\label{eq:107}
\end{equation}

Equations (\ref{eq:106}) and (\ref{eq:107}) clearly display the
fundamental difference between the volume and shear components of
the creep deformation. To simplify the discussion, suppose the material
is at phase-field equilibrium, $\Phi_{m}^{*}=0$. Then, by Eq.(\ref{eq:106})
the site generation (and thus volume creep) ceases when the driving
force $(\omega-\tilde{\sigma}_{h})$ turns to zero. The material reaches
equilibrium with respect to the \emph{net} production and annihilation
of sites. By contrast, Eq.(\ref{eq:107}) shows that the shear creep
\emph{never} stops as long as a shear stress exists in the material.
If $\mathcal{S}_{rr}\neq0$, the material continues to shear until
it reaches a hydrostatic state of stress (if this is permitted by
the boundary conditions). As indicated earlier, this type of shear
flow could occur, e.g., by the growth and dissolution of crystal planes
with different crystallographic orientations while preserving the
net number of sites. If this mechanism cannot operate, we have $\mathcal{S}_{rr}=0$
and the material is only capable of isotropic site generation causing
volume expansion or contraction. As already indicated, the tensor
character of the creep deformation rate and its splitting into the
volume and shear components was identified by Svoboda \emph{et al.}\cite{Svoboda2006,Fischer2011}

Eqs.(\ref{eq:104})-(\ref{eq:107}) also demonstrate that for an isotropic
material, diffusion is decoupled from creep deformation in the sense
of irreversible thermodynamics. Diffusion can offer a mechanism of
creep (hence the term ``diffusional'' creep) and may (or may not)
kinetically control the total deformation rate. However, diffusion
fluxes alone cannot \emph{cause} creep deformation and creep deformation
cannot \emph{cause} diffusion fluxes.

\subsubsection{Example of phenomenological relations for anisotropic materials\label{sub:anisotropic}}

The above equations rely on the assumption that the material is isotropic.
While this assumptions is adequate for fluids, polycrystalline materials
can possess a lower symmetry due to the crystallinity of the grains,
orientational texture or certain features of the microstructure. In
such cases, the form of the phenomenological equations is established
by analyzing the effects of the symmetry operations available in the
particular material on the individual terms in the entropy production.
Symmetry operations perform differently on fluxes and forces of different
tensor character. Thus, the tensor-split form of the entropy production
given by Eq.(\ref{eq:89}) can be taken as the starting point for
this analysis. A detailed analysis of anisotropic materials is beyond
the scope of this paper and we will restrict the discussion to one
example.

In simple cases the symmetry restrictions can be understood without
resorting to rigorous analysis. For example, suppose the only mechanism
of site generation and annihilation is the growth or dissolution of
crystal planes normal to a certain crystallographic direction defined
by a unit normal $\mathbf{n}$. All other properties of the material
related to diffusion and phase fields are assumed to remain fully
isotropic. Retracing the derivation of the entropy production for
this particular case, the site generation term becomes $(\tilde{\sigma}_{n}-\omega)R_{n}$,
where $\tilde{\sigma}_{n}=\mathbf{n}\cdot\tilde{\mathbf{\boldsymbol{\sigma}}}\cdot\mathbf{n}$
is the normal stress on the growing or dissolving crystal planes and
$R_{n}=\mathbf{n}\cdot\mathbf{R}\cdot\mathbf{n}$ is the normal creep
rate (rate of permanent tension-compression parallel to $\mathbf{n}$). 

In this case, the site generation is represented by only the product
of the scalar ``flux'' $R_{n}$ and the scalar force $(\omega-\tilde{\sigma}_{n})$.
As such, this term belongs to the first line in Eq.(\ref{eq:89})
and can couple to the scalar equations for the phase-field evolution.
The diffusion equations (\ref{eq:104}) remain unchanged but Eqs.(\ref{eq:105})-(\ref{eq:107})
are replaced by
\begin{eqnarray}
\dfrac{d^{L}\varphi_{k}}{dt} & = & -\dfrac{1}{T}\sum_{m=1}^{K}B_{km}\Phi_{m}^{*}-\dfrac{1}{T}B_{kr}\left(\omega-\tilde{\sigma}_{n}\right),\enskip\enskip k=1,...,K\label{eq:108}\\
R_{n} & = & -\dfrac{1}{T}\sum_{m=1}^{K}B_{rm}\Phi_{m}^{*}-\dfrac{1}{T}B_{rr}\left(\omega-\tilde{\sigma}_{n}\right).\label{eq:109}
\end{eqnarray}
Note that the separate shear creep equation (\ref{eq:107}) is now
redundant while the volume creep driven by $(\omega-\tilde{\sigma}_{h})$
has been replaced by uniaxial tension-compression creep driven by
$(\omega-\tilde{\sigma}_{n})$.

\section{Examples of application \label{sec:Simple-model}}

\subsection{Model formulation}

To illustrate the theory we will apply it to a simple one-dimensional
system. Namely, we consider an elemental bicrystal with a symmetrical
(e.g., {[}001{]} twist) GB. The grains are treated as isotropic media
and the entire bicrystal is assumed to possess the axial symmetry
($\infty/m$) around the coordinate axis $x_{1}$ normal to the GB
plane. The system is characterized by a single phase field $\varphi$
with the far-field values $\varphi=0$ in one grain and $\varphi=1$
in the other. This field can be interpreted, e.g., as the angle of
lattice rotation around the $x_{1}$ axis normalized by the total
lattice misorientation angle between the grains.

The lattice supports vacancies but not interstitials. Vacancies can
be generated only within the GB region and only by the growth or dissolution
of lattice planes parallel to the GB. We neglect thermal expansion
and the effect of vacancies on the lattice parameter. Thus, the latter
can only be altered by elastic strains. 

Elastic deformation of the lattice is described by the isotropic linear
elasticity theory with Hooke's law
\begin{equation}
\varepsilon_{ij}=\dfrac{1+\nu}{E}\sigma_{ij}-3\dfrac{\nu}{E}\delta_{ij}\sigma_{h},\label{eq:120}
\end{equation}
where $\varepsilon_{ij}$ is the small-strain tensor, $E$ is the
Young modulus and $\nu$ is Poisson's ratio. Both $E$ and $\nu$
are considered constant. Although we use the small-strain approximation
for elasticity, the total deformation of the material is allowed to
be finite due to the creep process. Since the deformations are assumed
to be slow and the viscous energy dissipation is neglected, the dynamic
stress is identical to the static. The classical part of the stress
will be denoted $\sigma_{ij}$ without the tilde. 

Due to the axial symmetry of the problem, the stress components $\sigma_{11}$
(normal stress) and $\sigma_{22}=\sigma_{33}$ (lateral stresses)
depend only on the distance $x$ along the $x_{1}$ axis, the shear
components being zero. Likewise, the normal strain component $\varepsilon_{11}$
is a function of $x$, the lateral strains $\varepsilon_{22}=\varepsilon_{33}$
are assumed to be fixed, and the shear strains are zero. Under these
conditions, knowing only the function $\sigma_{11}(x)$ and using
Hooke's law one can recover 
\begin{equation}
\sigma_{22}(x)=\sigma_{33}(x)=\dfrac{\nu}{1-\nu}\sigma_{11}(x)+\dfrac{E}{1-\nu}\varepsilon_{22}\label{eq:121}
\end{equation}
and
\begin{equation}
\varepsilon_{11}(x)=\dfrac{(1+\nu)(1-2\nu)}{E(1-\nu)}\sigma_{11}(x)-\dfrac{2\nu}{(1-\nu)}\varepsilon_{22}.\label{eq:123}
\end{equation}
The volume per site is
\begin{equation}
\Omega=\Omega_{0}\left(1+K_{T}\sigma_{h}\right),\label{eq:124}
\end{equation}
where $\Omega_{0}$ is the stress-free value of $\Omega$, $K_{T}=3(1-2\nu)/E$
is the isothermal compressibility and 
\begin{equation}
\sigma_{h}(x)=\dfrac{(1+\nu)}{(1-\nu)}\sigma_{11}(x)+\dfrac{2E}{(1-\nu)}\varepsilon_{22}\label{eq:125}
\end{equation}
is the hydrostatic part of the stress tensor. 

To describe thermodynamics of the solid, two adjustments will be made
with respect to the previous discussion. First, for practical convenience
all thermodynamic properties will be described in terms of the Helmholtz
free energy instead of the internal energy. All previous expressions
for the entropy production remain unchanged, except that the derivatives
of the internal energy density (e.g., $\partial u/\nabla\varphi$)
taken previously at a fixed entropy are replaced by derivatives of
the free energy density $f$ (e.g., $\partial f/\nabla\varphi$) is
taken at a fixed temperature.%
\footnote{This becomes clear by applying the Legendre transformation with respect
to $s$ in Eq.(\ref{eq:17-3}), which becomes $df=-sdT+...\textrm{(remaining\:\ terms)}$.
The differential coefficients in the remaining terms are now partial
derivatives of $f$ at constant $T$ instead of the derivatives of
$u$ at constant $s$.%
} Secondly, the fundamental equation for a specific material usually
comes from statistical-mechanical models and is formulated in terms
of the site fractions of the components and thermodynamic properties
(e.g., free energy) per site. We will therefore use the site fractions
of atoms $c$ and vacancies $c_{v}$, keeping in mind that only one
of them can be used as an independent variable ($c+c_{v}=1$). It
is implicit in this treatment that the GB structure is composed of
sites and can be obtained by an appropriate distortion of the lattice.

We postulate the fundamental equation of the solid in the form
\begin{equation}
f\left(T,c_{v},\boldsymbol{\varepsilon}\right)=\dfrac{1}{\Omega}\left[gc+g_{v}c_{v}+kT\left(c\ln c+c_{v}\ln c_{v}\right)\right]+w\left(\varphi\right)+\dfrac{1}{2}\epsilon\left(\nabla\varphi\right)^{2}+e\left(\boldsymbol{\varepsilon}\right).\label{eq:126}
\end{equation}
Here, $g$ and $g_{v}$ are parameters of the ideal solution model
for atoms and vacancies, $k$ is Boltzmann's factor, 
\begin{equation}
w\left(\varphi\right)=W\varphi^{2}\left(1-\varphi\right)^{2}\label{eq:127}
\end{equation}
is a double-well function with an amplitude $W$ creating a free energy
barrier between the two lattice orientations, $\epsilon$ is the gradient
energy coefficient\cite{Cahn58a} considered constant, and finally
$e\left(\boldsymbol{\varepsilon}\right)$ is the elastic strain energy
density of the lattice. The latter is quadratic in strains (and thus
stresses) and will not be detailed here since this term will be neglected.
The expression in the square brackets is the free energy of a uniform
ideal solution per site. Note that this solution is treated classically,
i.e., without a gradient term in $c$. By the symmetry of the problem,
the gradient $\nabla\varphi$ has only one nonzero component $\nabla_{x}\varphi$.

This model is different from previous non-classical interface models
with elasticity. Rottman \cite{Rottman1988} proposed a Landau theory
of coherent phase boundaries and computed the interface stress and
other excess properties by including a gradient term in strains. Johnson
\cite{Johnson2000} modeled a phase boundary between two binary substitutional
solutions using a gradient term in composition. His model includes
a compositional strain and, by contrast to Rottman's work,\cite{Rottman1988}
treats the elastic strain energy purely classically. Johnson carefully
derives integral expressions for the interface free energy, interface
stress and interface strain. While these workers were focused on the
equilibrium state of the interface, Levitas \cite{Levitas2013} recently
proposed a time-dependent model with a single non-classical order
parameter $\varphi$ and elastic strain energy. Assuming mechanical
equilibrium, he solved the phase-field evolution equation of the form
$\partial\varphi/\partial t=-L\Phi^{*}$ and studied in detail the
dynamics of the interface stress at the non-equilibrium interface.
His model does not include diffusion or site generation.

The subsequent calculations will be limited to first order in stresses
and strains. Thus, the elastic energy strain term appearing Eq.(\ref{eq:126})
and propagating to all other equations will be neglected. This approximation
is sufficient for demonstrating some simple results of the model.

From Eq.(\ref{eq:126}) we obtain the diffusion potential $M$ of
atoms relative vacancies, 
\begin{equation}
M=g-g_{v}+kT\ln\dfrac{c}{c_{v}},\label{eq:128}
\end{equation}
and thus the grand potential density 
\begin{equation}
\omega=f-M\dfrac{c}{\Omega}=\dfrac{1}{\Omega}\left(g_{v}+kT\ln c_{v}\right)+w\left(\varphi\right)+\dfrac{1}{2}\epsilon\left(\nabla_{x}\varphi\right)^{2}.\label{eq:129}
\end{equation}
The variational derivative of $f$ with respect to the phase field
is given by the usual expression
\begin{equation}
\Phi^{*}=\dfrac{\partial f}{\partial\varphi}-\nabla_{x}\cdot\dfrac{\partial f}{\partial\nabla_{x}\varphi}=w^{\prime}\left(\varphi\right)-\epsilon\nabla_{x}^{2}\varphi.\label{eq:130}
\end{equation}
Finally, the non-classical tensor $\mathbf{A}^{*}$ defined by Eq.(\ref{eq:23-1-1})
is $\mathbf{A}^{*}=-\epsilon\nabla\varphi\nabla\varphi$ and has only
one nonzero component

\begin{equation}
A_{11}^{*}=-\epsilon\left(\nabla_{x}\varphi\right)^{2}.\label{eq:134}
\end{equation}

\subsection{The state of equilibrium}

Before discussing the dynamics of creep deformation, we will find
the state of thermodynamic equilibrium of the system. We assume that
the system is already in thermal equilibrium and thus the temperature
is uniform. The phase-field equilibrium condition $\Phi^{*}=0$ reduces
to the standard equation\cite{Cahn58a} 
\begin{equation}
w\left(\varphi\right)=\dfrac{1}{2}\epsilon\left(\nabla_{x}\varphi\right)^{2}\label{eq:131}
\end{equation}
predicting the phase-field profile
\begin{equation}
\varphi(x)=\dfrac{1}{2}-\dfrac{1}{2}\tanh\dfrac{x}{2\sqrt{\epsilon/2W}}.\label{eq:132}
\end{equation}
Using Eq.(\ref{eq:131}), the grand potential density (\ref{eq:129})
becomes
\begin{equation}
\omega=\dfrac{1}{\Omega}\left(g_{v}+kT\ln c_{v}\right)+\epsilon\left(\nabla_{x}\varphi\right)^{2}.\label{eq:133}
\end{equation}

The mechanical equilibrium condition (\ref{eq:75}) reduces to $\sigma_{11}+A_{11}^{*}=\textrm{const}$,
giving 
\begin{equation}
\sigma_{11}(x)=\sigma_{11}^{\infty}+\epsilon\left(\nabla_{x}\varphi\right)^{2},\label{eq:135}
\end{equation}
where $\sigma_{11}^{\infty}$ is the coordinate-independent normal
stress inside the grains. The site-generation equilibrium condition
is $\omega-\sigma_{11}=0$ (Sect.~\ref{sub:anisotropic}). 

Using the above equations we have
\begin{equation}
g_{v}+kT\ln c_{v}=\sigma_{11}^{\infty}\Omega,\label{eq:136}
\end{equation}
which can be rewritten
\begin{equation}
kT\ln\dfrac{c_{v}}{c_{v}^{0}}=\sigma_{11}^{\infty}\Omega,\label{eq:137}
\end{equation}
where $c_{v}^{0}$ is the equilibrium vacancy concentration in the
absence of normal stress. The obtained Eq.(\ref{eq:137}) reproduces
Herring's relation for the effect of stresses on the vacancy concentration
in solids.\cite{herring49,Herring1950} 

Using Eq.(\ref{eq:135}), the equilibrium grand-potential density
across the GB becomes
\begin{equation}
\omega(x)=\sigma_{11}^{\infty}+\epsilon\left(\nabla_{x}\varphi\right)^{2}\label{eq:138}
\end{equation}
with $\omega^{\infty}=\sigma_{11}^{\infty}$ inside the grains.

The GB free energy $\gamma$ is computed as the excess of $\omega$
over the homogeneous grains:
\begin{equation}
\gamma=\intop_{-\infty}^{\infty}\left[\omega(x)-\omega^{\infty}\right]dx=\epsilon\intop_{-\infty}^{\infty}\left(\nabla_{x}\varphi\right)^{2}dx=\sqrt{\dfrac{\epsilon W}{18}}.\label{eq:139}
\end{equation}
The interface stress of the GB is isotropic, $\tau_{22}=\tau_{33}\equiv\tau$,
and is computed as the excess of $\sigma_{22}$. Using Eqs.(\ref{eq:121})
and (\ref{eq:135}),
\begin{equation}
\sigma_{22}(x)=\dfrac{\nu}{1-\nu}\sigma_{11}^{\infty}+\dfrac{E}{1-\nu}\varepsilon_{22}+\epsilon\dfrac{\nu}{1-\nu}\left(\nabla_{x}\varphi\right)^{2},\label{eq:140}
\end{equation}
where only the last term contributes to the excess. Thus,
\begin{equation}
\tau=\intop_{-\infty}^{\infty}\left[\sigma_{22}(x)-\sigma_{22}^{\infty}\right]dx=\epsilon\dfrac{\nu}{1-\nu}\intop_{-\infty}^{\infty}\left(\nabla_{x}\varphi\right)^{2}dx=\dfrac{\nu}{1-\nu}\sqrt{\dfrac{\epsilon W}{18}}.\label{eq:141}
\end{equation}
We see that in this particular model $\gamma$ and $\tau$ are proportional
to each other and independent of the stressed state of the grains.
They are generally different unless the materials is incompressible
($\nu=1/2$).

We can also compute the GB excess volume $E_{11}$ per unit area as
the excess of the strain component $\varepsilon_{11}$. Using Eq.(\ref{eq:123}),

\begin{equation}
\varepsilon_{11}(x)=\dfrac{(1+\nu)(1-2\nu)}{E(1-\nu)}\sigma_{11}^{\infty}-\dfrac{2\nu}{(1-\nu)}\varepsilon_{22}+\epsilon\dfrac{(1+\nu)(1-2\nu)}{E(1-\nu)}\left(\nabla_{x}\varphi\right)^{2},\label{eq:142}
\end{equation}
where only the last term contributes to the excess. Thus,
\begin{equation}
E_{11}=\epsilon\dfrac{(1+\nu)(1-2\nu)}{E(1-\nu)}\intop_{-\infty}^{\infty}\left(\nabla_{x}\varphi\right)^{2}dx=\dfrac{(1+\nu)(1-2\nu)}{E(1-\nu)}\gamma,\label{eq:143}
\end{equation}
where we used Eq.(\ref{eq:139}). In this model the GB excess volume
is proportional to the GB free energy. For an incompressible material
($\nu=1/2$) we correctly obtain $E_{11}=0$.

\subsection{Dynamics of creep }

\subsubsection{Dynamic equations}

We now consider irreversible processes involving vacancy diffusion,
site generation and GB motion. Due to the simplified geometry of this
example we will obviously not be able to model a real three-dimensional
creep process taking place in polycrystalline materials. However,
several elementary steps of this  process can be reproduced and studied.

The dynamic equations of the system are based on Eqs.(\ref{eq:104}),
(\ref{eq:108}) and (\ref{eq:109}) adapted to this model. Neglecting
all cross-effects we have
\begin{equation}
J_{x}^{L}=-\dfrac{L}{T}\nabla_{x}M,\label{eq:145}
\end{equation}
\begin{equation}
\dfrac{\partial\varphi}{\partial t}+v_{L}\nabla_{x}\varphi=-\dfrac{B}{T}\left[w^{\prime}\left(\varphi\right)-\epsilon\nabla_{x}^{2}\varphi\right],\label{eq:146}
\end{equation}
\begin{equation}
R_{n}=-\dfrac{B_{rr}}{T}\left(\omega-\sigma_{11}\right)\label{eq:147}
\end{equation}
with three kinetic coefficients $L$, $B$ and $B_{rr}$. Here $v_{L}$
is the lattice velocity and $R_{n}$ is the creep deformation rate
(i.e., rate of the sample elongation or compression) in the $x$-direction.
In keeping with the first-order approximation in stress adopted here,
we replace the elastically deformed site volume $\Omega$ by its stress-free
value $\Omega_{0}$. In addition, $R_{n}=\nabla_{x}v_{L}-\dot{\varepsilon}_{11}$
($\dot{\varepsilon}_{11}$ being the elastic tensile strain rate)
will be approximated by simply $\nabla_{x}v_{L}$. This approximation
is applicable to steady-state creep under a sustained load when the
elastic deformation does not practically change with time while the
permanent deformation due to creep increases and may reach tens of
per cent. In this regime, this approximation should work. Finally,
we assume that the system maintains mechanical equilibrium at all
times and thus Eq.(\ref{eq:135}) remains satisfied. 

The diffusion equation (\ref{eq:145}) can be conveniently reformulated
in terms of the vacancy flux $J_{xv}^{L}=-J_{x}^{L}$ and the vacancy
site fraction $c_{v}$. Taking into account that $c_{v}\ll1$ we have
\begin{equation}
J_{xv}^{L}=-D_{v}\nabla_{x}\dfrac{c_{v}}{\Omega_{0}},\label{eq:148}
\end{equation}
where $D_{v}=k\Omega_{0}L/c_{v}$ is the vacancy diffusion coefficient
assumed to be constant. Rewriting also the continuity equation (\ref{eq:5})
in terms of $c_{v}$ we finally obtain the vacancy diffusion equation
\begin{equation}
\dfrac{\partial c_{v}}{\partial t}+v_{L}\nabla_{x}c_{v}-D_{v}\nabla_{x}^{2}c_{v}=\nabla_{x}v_{L}.\label{eq:149}
\end{equation}

The driving force for site generation is $(\omega-\sigma_{11})$ which
by Eqs.(\ref{eq:129}) and (\ref{eq:135}) equals 
\begin{equation}
\dfrac{kT}{\Omega_{0}}\ln\dfrac{c_{v}}{c_{v}^{0}}-\sigma_{11}^{\infty}+w\left(\varphi\right)-\dfrac{1}{2}\epsilon\left(\nabla_{x}\varphi\right)^{2}.\label{eq:144}
\end{equation}
The kinetic coefficient controlling the site generation is postulated
in the form
\begin{equation}
\dfrac{B_{rr}}{T}=B_{r}w\left(\varphi\right)\label{eq:150}
\end{equation}
where $B_{r}$ is a constant. This form ensures that site generation
occurs only within the GB region and not inside the grains where $w\left(\varphi\right)\ll1$.
Thus, the site generation equation (\ref{eq:147}) becomes
\begin{equation}
\nabla_{x}v_{L}=-B_{r}w\left(\varphi\right)\left[\dfrac{kT}{\Omega_{0}}\ln\dfrac{c_{v}}{c_{v}^{0}}-\sigma_{11}^{\infty}+w\left(\varphi\right)-\dfrac{1}{2}\epsilon\left(\nabla_{x}\varphi\right)^{2}\right].\label{eq:151}
\end{equation}

The three equations (\ref{eq:146}), (\ref{eq:149}) and (\ref{eq:151})
with appropriate initial and boundary conditions describe the entire
dynamics of our system.

\subsubsection{Numerical examples}

For numerical calculations it is convenient to non-dimensionalize
the above equations. We introduce the dimensionless time $\tau=2WD_{v}t/\epsilon$,
dimensionless coordinate $\xi=x\sqrt{2W/\epsilon}$, dimensionless
lattice velocity $\eta=\partial\xi_{L}/\partial\tau=(\sqrt{\epsilon/2W}/D_{v})v_{L}$
and normalized vacancy concentration $\zeta=c_{v}/c_{v}^{0}$. In
terms of these variables, the equilibrium interface thickness is approximately
$\Delta\xi\sim1$ and the diffusion time across the interface is approximately
$\Delta\tau\sim1$. The dynamic equations to be solved take the form
\begin{equation}
\dfrac{\partial\varphi}{\partial\tau}=-\eta\dfrac{\partial\varphi}{\partial\xi}-\beta_{\varphi}\left[\varphi\left(2\varphi^{2}-3\varphi+1\right)-\dfrac{\partial^{2}\varphi}{\partial\xi^{2}}\right],\label{eq:152}
\end{equation}
\begin{equation}
\dfrac{\partial\zeta}{\partial\tau}=-\eta\dfrac{\partial\zeta}{\partial\xi}+\dfrac{\partial^{2}\zeta}{\partial\xi^{2}}+\dfrac{1}{c_{v}^{0}}\dfrac{\partial\eta}{\partial\xi},\label{eq:153}
\end{equation}
\begin{equation}
\dfrac{\partial\eta}{\partial\xi}=-\beta_{s}\varphi^{2}\left(1-\varphi\right)^{2}\left[\ln\zeta-a_{\sigma}+a_{w}\left(\varphi^{2}\left(1-\varphi\right)^{2}-\left(\dfrac{\partial\varphi}{\partial\xi}\right)^{2}\right)\right].\label{eq:154}
\end{equation}
Here,
\begin{equation}
\beta_{\varphi}=\dfrac{B\epsilon}{D_{v}T}\label{eq:155}
\end{equation}
and
\begin{equation}
\beta_{s}=\dfrac{B_{r}\epsilon kT}{2\Omega_{0}D_{v}}\label{eq:156}
\end{equation}
are dimensionless kinetic coefficients characterizing the rates of
the phase-field evolution and site generation, respectively, relative
to diffusion. The two other dimensionless parameters, $a_{\sigma}=\sigma_{11}^{\infty}\Omega_{0}/kT$
and $a_{w}=W\Omega_{0}/kT$, characterize the strength of the applied
stress and the phase-field barrier, respectively, relative to the
thermal energy $kT$.

The system of equations (\ref{eq:152})-(\ref{eq:154}) was solved
numerically on an interval $0\leq\xi\leq\mathcal{L}$. The GB was
initially placed at $\xi=\mathcal{L}/2$ by solving Eq.(\ref{eq:152})
with $\eta\equiv0$ and the boundary conditions 
\begin{equation}
\varphi(0,\tau)=0,\enskip\enskip\varphi(\mathcal{L},\tau)=1.\label{eq:157}
\end{equation}
The obtained phase-field profile was very close to the infinite-system
solution (\ref{eq:132}). The boundary conditions (\ref{eq:157})
were maintained throughout the subsequent calculations. The equilibrium
vacancy concentration was chosen to be $c_{v}^{0}=10^{-4}$. This
is an order of magnitude larger than typical experimental values at
the melting point of metals. However, the choice was dictated by computational
efficiency and was deemed to be sufficient for qualitative demonstration
of the effects.

For the velocity field $\eta(\xi,\tau)$ we used the initial condition
$\eta(\xi,0)=0$ and the boundary condition $\eta(0,\tau)=0$ which
fixes the position of the left end of the left grain. For the vacancy
concentration field $\zeta(\xi,\tau)$ we used different initial conditions
as specified below. Under these boundary conditions the system is
open at its right end ($\xi=\mathcal{L}$) where the atoms as well
as crystal planes are allowed to enter or leave the system. 

\begin{figure}
\noindent \begin{centering}
\textbf{(a)}\enskip{}\includegraphics[scale=0.5]{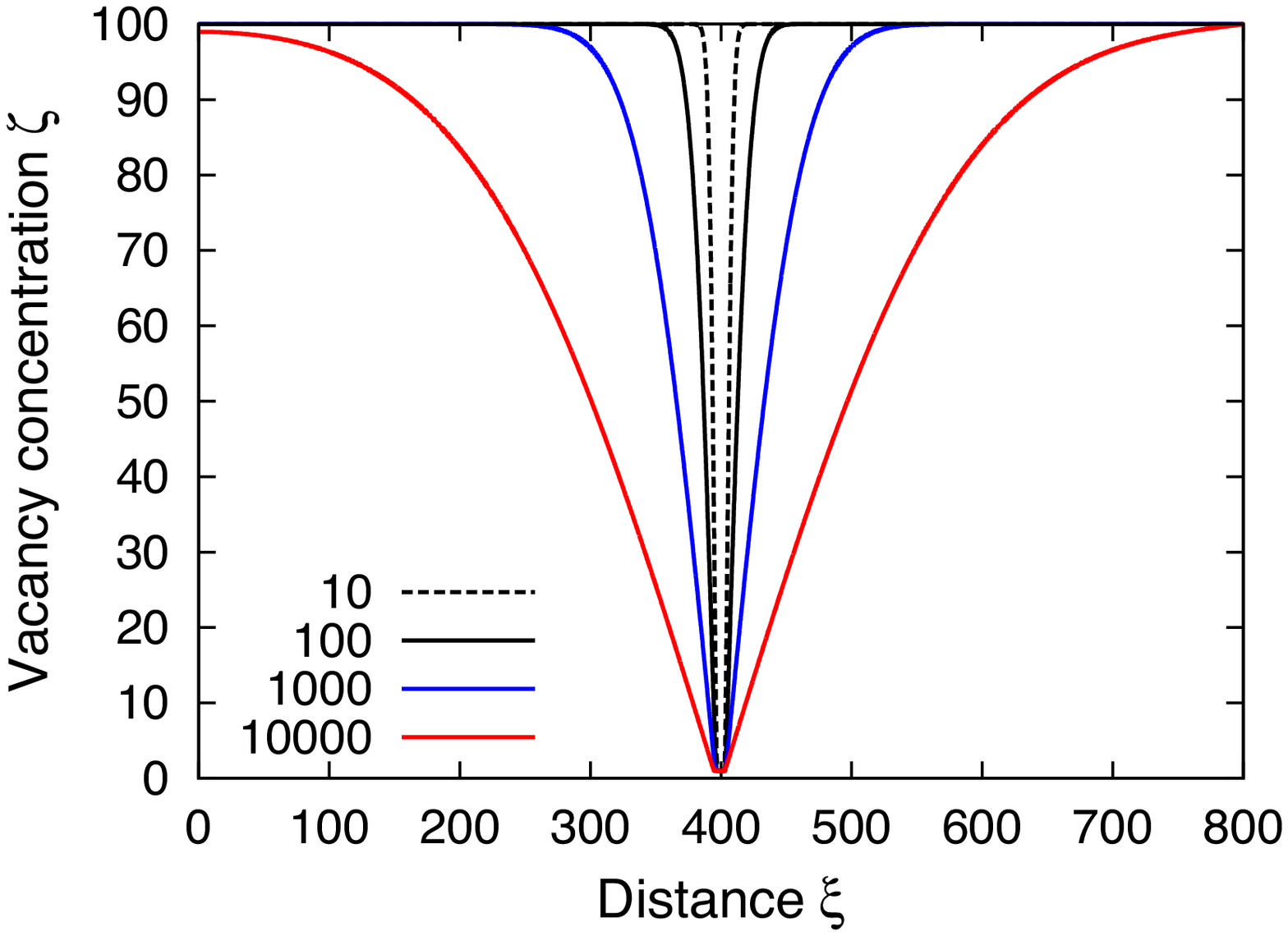}\vspace{0.12in}

\par\end{centering}

\noindent \begin{centering}
\textbf{(b)}\enskip{}\includegraphics[clip,scale=0.5]{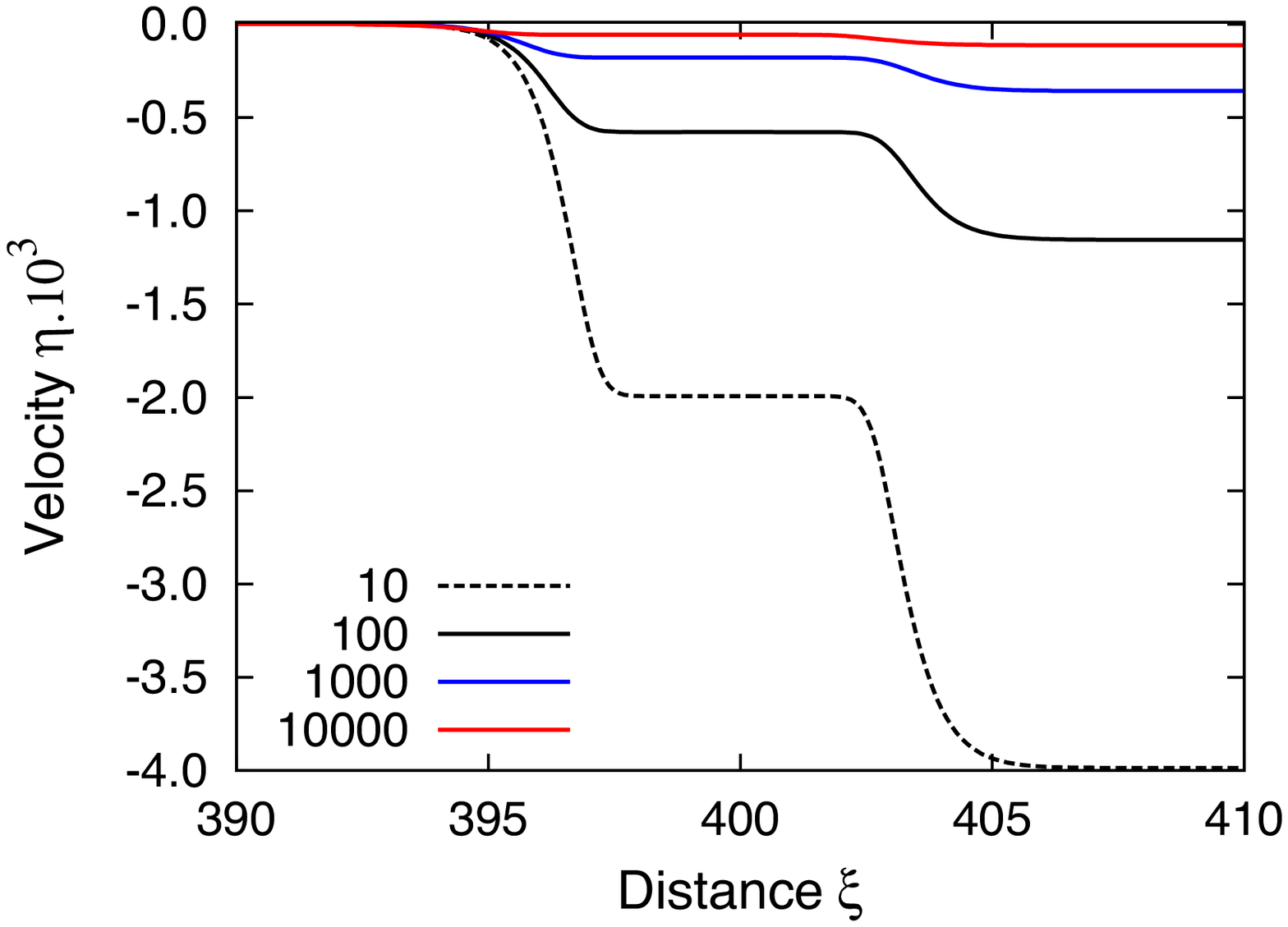}
\par\end{centering}

\noindent \begin{centering}
\vspace{0.1in}

\par\end{centering}

\noindent \begin{centering}
\textbf{(c)}\enskip{}\includegraphics[clip,scale=0.5]{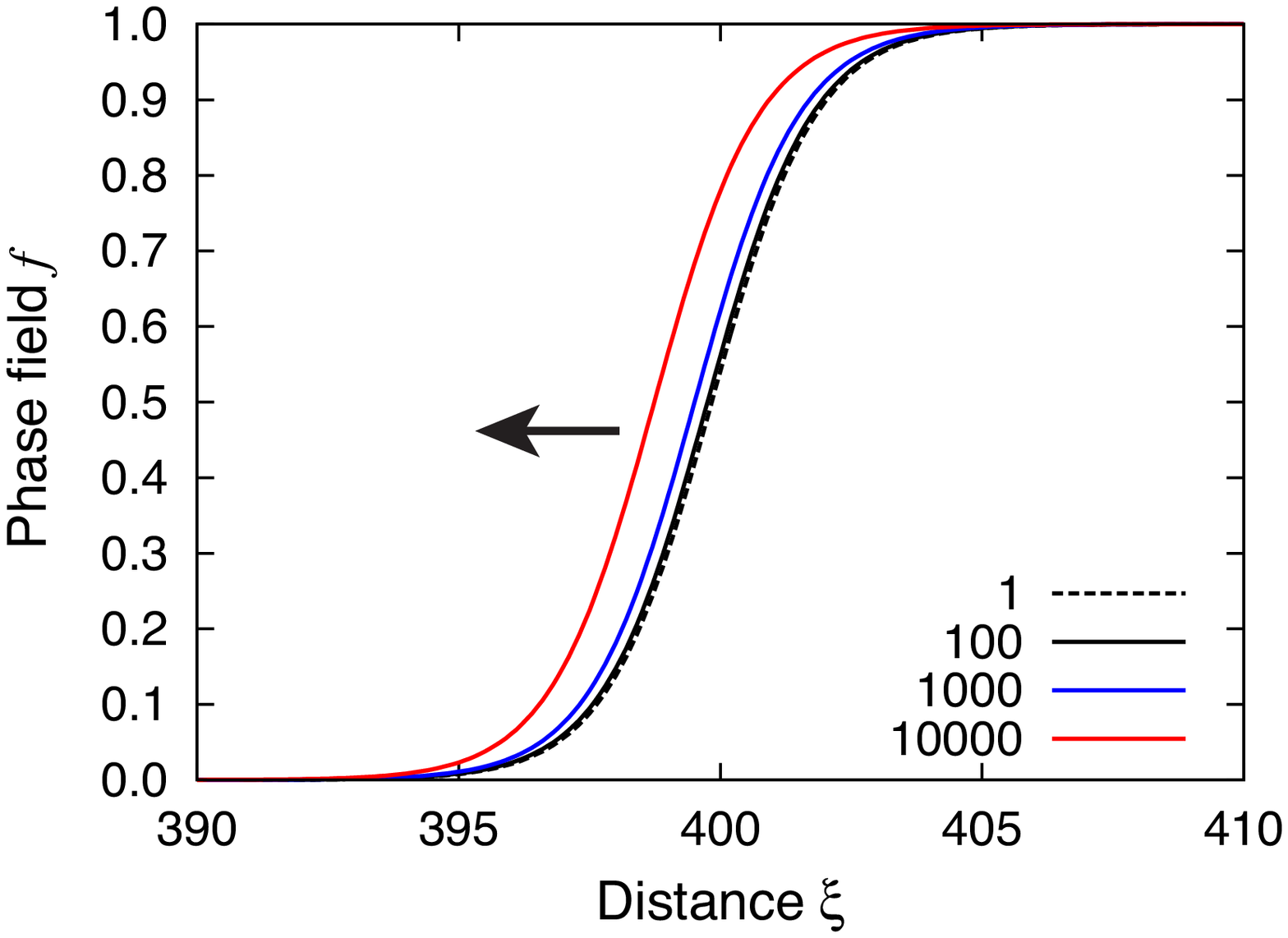}\vspace{0.12in}

\par\end{centering}

\caption{Dimensionless profiles of (a) vacancy concentration $\zeta$ (initial
value $\zeta=100$), (b) lattice velocity $\eta$ and (c) phase field
$\varphi$. The time $\tau$ is indicated in the legends. The model
parameters are $\beta_{\varphi}=1$, $\beta_{s}=0.8$, $a_{\sigma}=0$
(no stress) and $a_{w}=0.5$. Note that the phase-field profile is
drifting to the left (indicated by the arrow) reflecting GB migration.\label{fig:Fig-2}}
\end{figure}

\emph{\uline{Example 1.}} We first consider a stress-free ($a_{\sigma}=0$)
bicrystal of length $\mathcal{L}=800$. The initial state is a uniform
vacancy over-saturation with concentration $\zeta=100$. We impose
a zero-flux condition $\partial\zeta/\partial\xi=0$ at the left end
($\mathcal{\xi}=0$) and a fixed-concentration condition $\zeta(\mathcal{L},\tau)=\zeta(\mathcal{L},0)$
at the right end. In the absence of the GB or when the latter is unable
to generate/eliminate sites ($\beta_{s}=0$), this initially uniform
concentration profile will not change with time. When $\beta_{s}>0$,
the GB starts to eliminate excess vacancies, creating a local concentration
minimum (Fig.~\ref{fig:Fig-2}). With time, this minimum deepens
and widens as the vacancy concentration in the GB reaches its equilibrium
value $\zeta=1$. This process is accompanied by elimination of crystal
planes in the GB region resulting in shortening of both grains and
thus a flow of the right grain to the left. This explains the uniform
negative velocity field on the right of the GB. The GB itself also
moves to the left, although slower than the right grain. Since the
vacancy concentration is small, vacancies from vast lattice volumes
must be absorbed to eliminate even a single lattice plane. It is not
surprising, therefore, that the GB displacement is much smaller than
the width of the vacancy diffusion zone around the boundary, which
eventually reaches the size of the sample. 

\begin{figure}
\noindent \begin{centering}
\textbf{(a)}\enskip{}\includegraphics[scale=0.5]{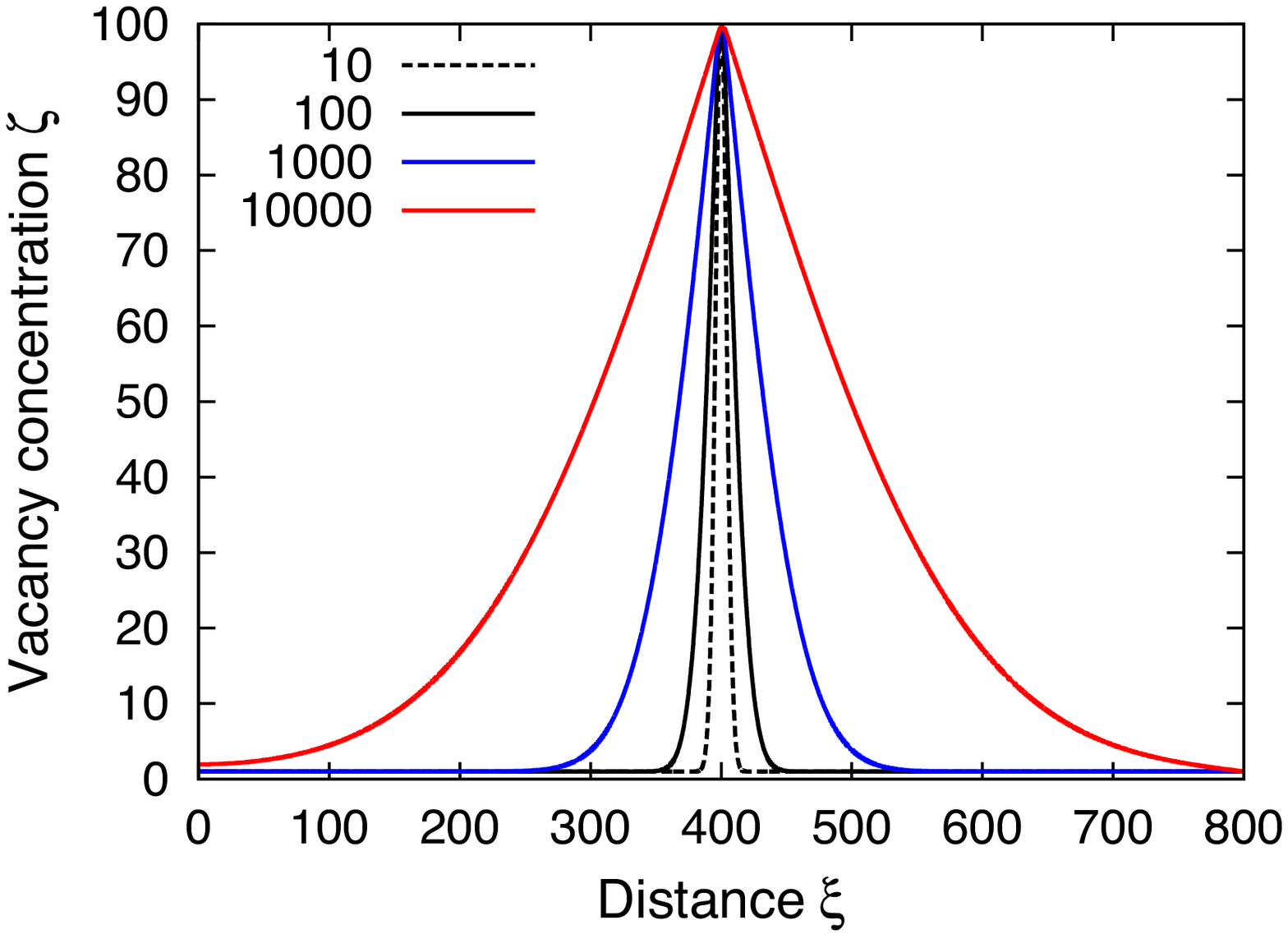}\vspace{0.12in}

\par\end{centering}

\noindent \begin{centering}
\textbf{(b)}\enskip{}\includegraphics[scale=0.5]{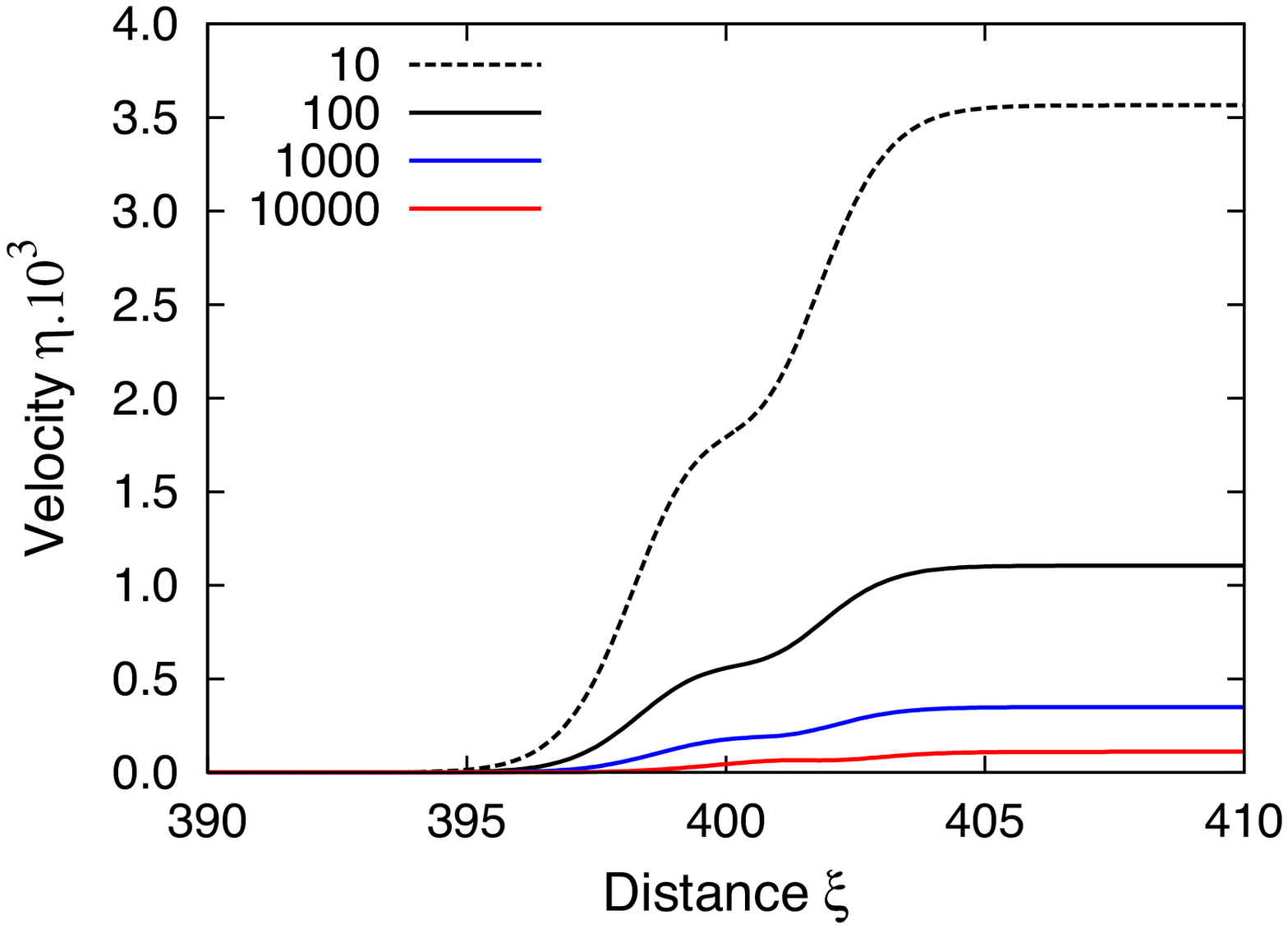}\vspace{0.12in}

\par\end{centering}

\noindent \begin{centering}
\textbf{(c)}\enskip{}\includegraphics[scale=0.5]{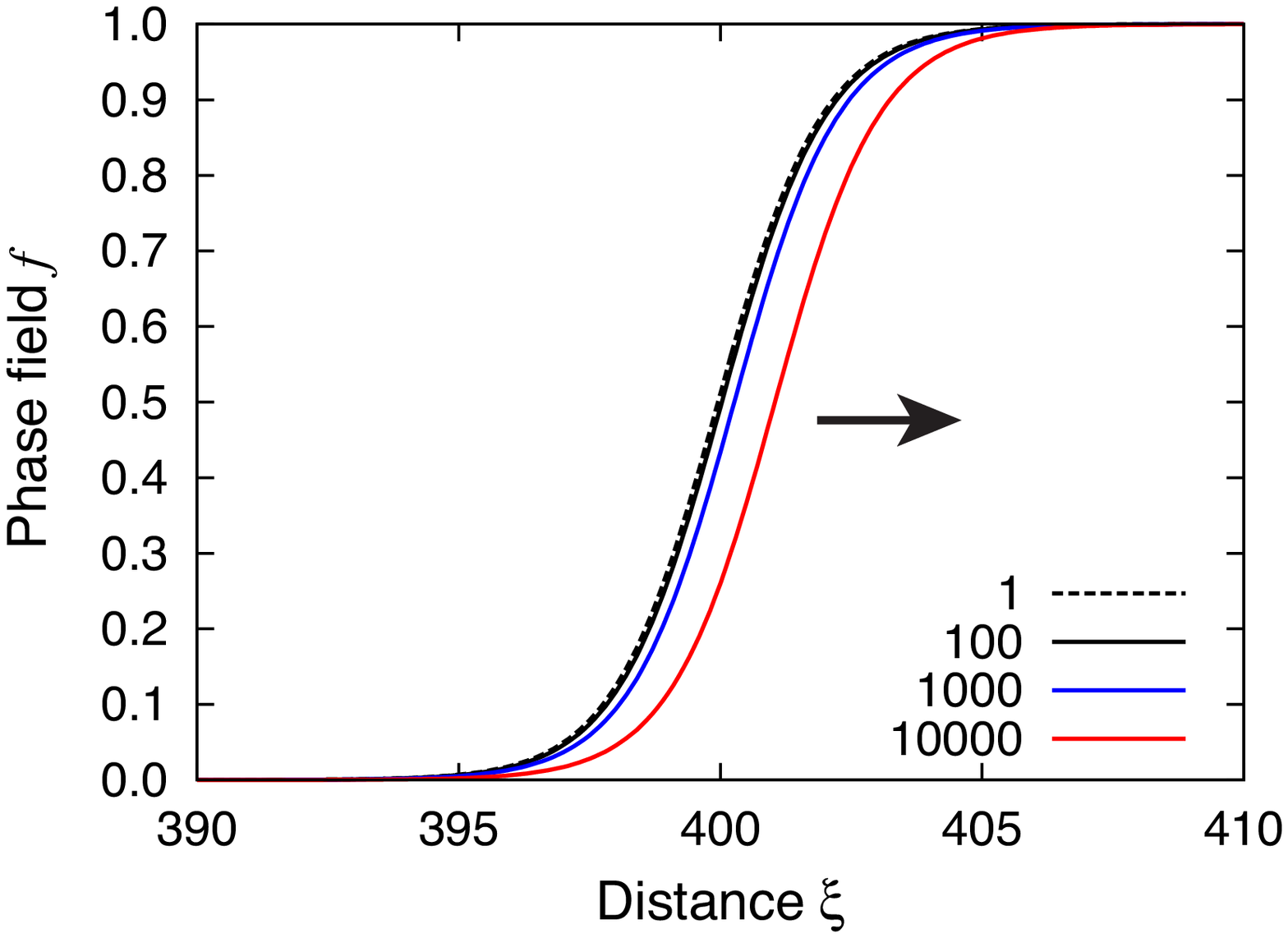}\vspace{0.12in}

\par\end{centering}

\caption{Dimensionless profiles of (a) vacancy concentration $\zeta$ (initial
value $\zeta=1$), (b) lattice velocity $\eta$ and (c) phase field
$\varphi$. The time $\tau$ is indicated in the legends. The model
parameters are $\beta_{\varphi}=1$, $\beta_{s}=0.8$, $a_{\sigma}=4.6$
(tensile stress) and $a_{w}=0.5$. Note that the phase-field profile
is drifting to the right (indicated by the arrow) reflecting GB migration.\label{fig:Fig-3}}
\end{figure}

\emph{\uline{Example 2.}} Next we consider the same bicrystal ($\mathcal{L}=800$)
subject to the same boundary conditions. Suppose it has been equilibrated
at zero value of the tensile stress. At a moment $\tau=0$ the stress
is suddenly raised to a value $a_{\sigma}=4.6$ (tension) corresponding
to the new equilibrium vacancy concentration $\zeta\approx99.5$.
To reach it, the GB generates vacancies producing a concentration
maximum that grows and widens with time (Fig.~\ref{fig:Fig-3}).
The vacancy generation occurs by embedding extra crystal planes on
either side of the GB, which results in the motion of the right grain
as well as the GB to the right. In this example, the application of
the tensile stress causes the growth of both grains by accretion of
material in the GB region, resulting in creep deformation of the sample.
As in the previous case, the GB displacement is small in comparison
with the width of the diffusion zone due to the small vacancy concentration.

\begin{figure}
\noindent \begin{centering}
\textbf{(a)}\enskip{}\includegraphics[scale=0.5]{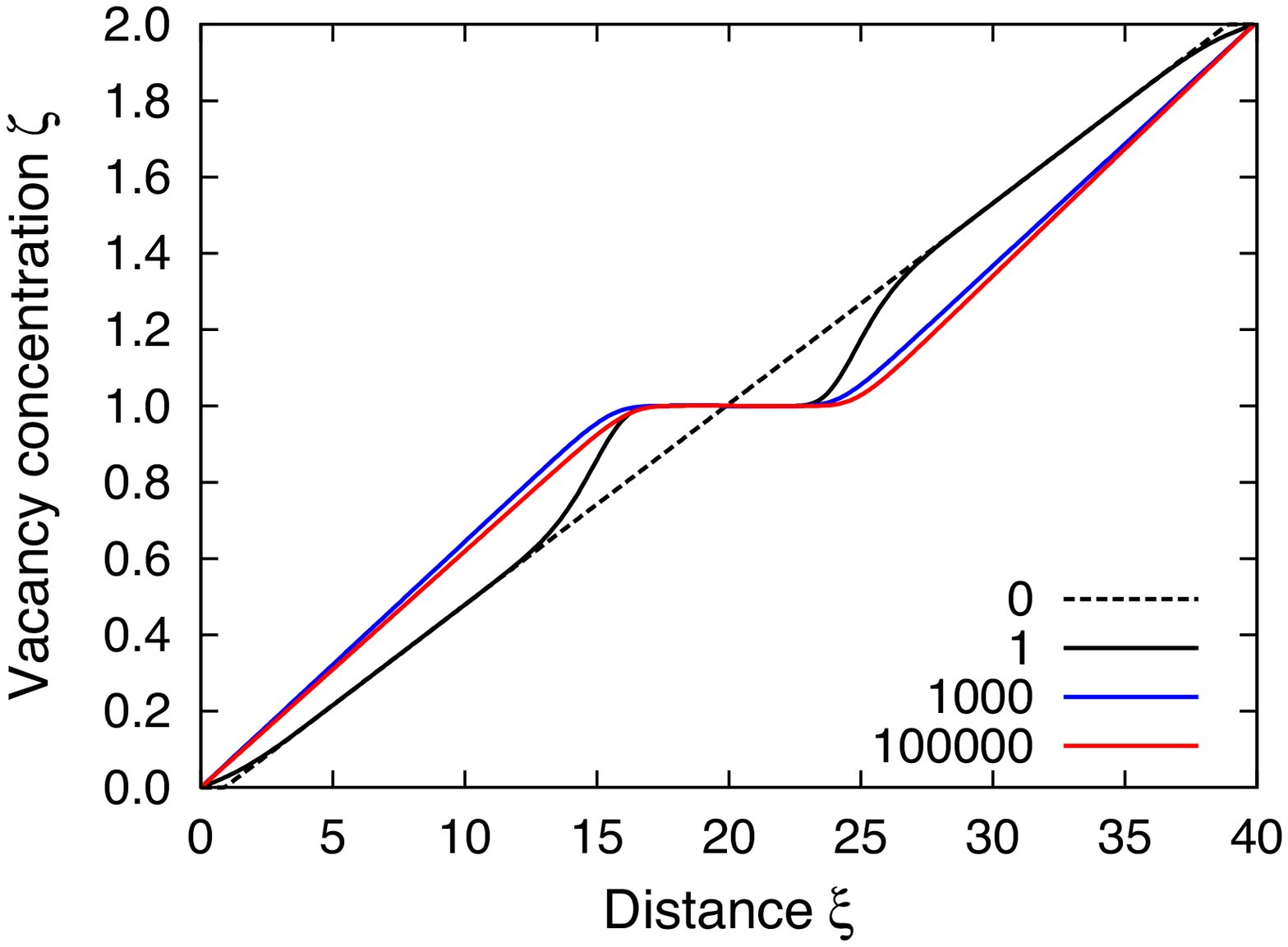}
\par\end{centering}

\noindent \begin{centering}
\vspace{0.12in}

\par\end{centering}

\noindent \begin{centering}
\textbf{(b)}\enskip{}\includegraphics[scale=0.5]{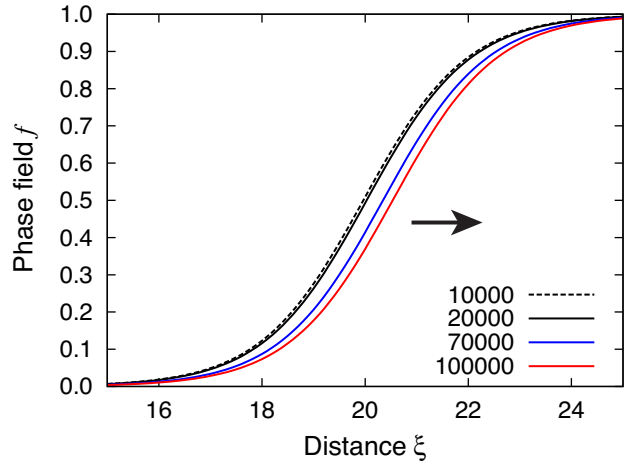}\vspace{0.12in}

\par\end{centering}

\caption{Dimensionless profiles of (a) vacancy concentration $\zeta$ and (b)
phase field $\varphi$. The time $\tau$ is indicated in the legends.
The model parameters are $\beta_{\varphi}=1$, $\beta_{s}=0.8$, $a_{\sigma}=0$
and $a_{w}=0.5$. The arrow indicates the trans-gradient induced GB
migration to the right.\label{fig:Fig-4}}
\end{figure}

\emph{\uline{Example 3.}} Suppose the bicrystal is stress-free
and a vacancy concentration gradient has been created around the initial
GB position. Computationally, this has been achieved by creating a
linear vacancy concentration profile increasing from $\zeta=0$ at
$\mathcal{\xi}=0$ to $\zeta=2$ at $\mathcal{\xi}=\mathcal{L}$ and
keeping these boundary values fixed (Fig.~\ref{fig:Fig-4}). To amplify
the concentration gradient, this calculation was performed in a smaller
system with $\mathcal{L}=40$. Note that in its initial position at
$\xi=\mathcal{L}/2$, the GB sees the equilibrium concentration $\zeta=1$.
Thus, this calculation is a test of the GB response to a vacancy concentration
gradient around the equilibrium value. 

\begin{figure}
\noindent \begin{centering}
\textbf{(a)}\enskip{}\includegraphics[scale=0.5]{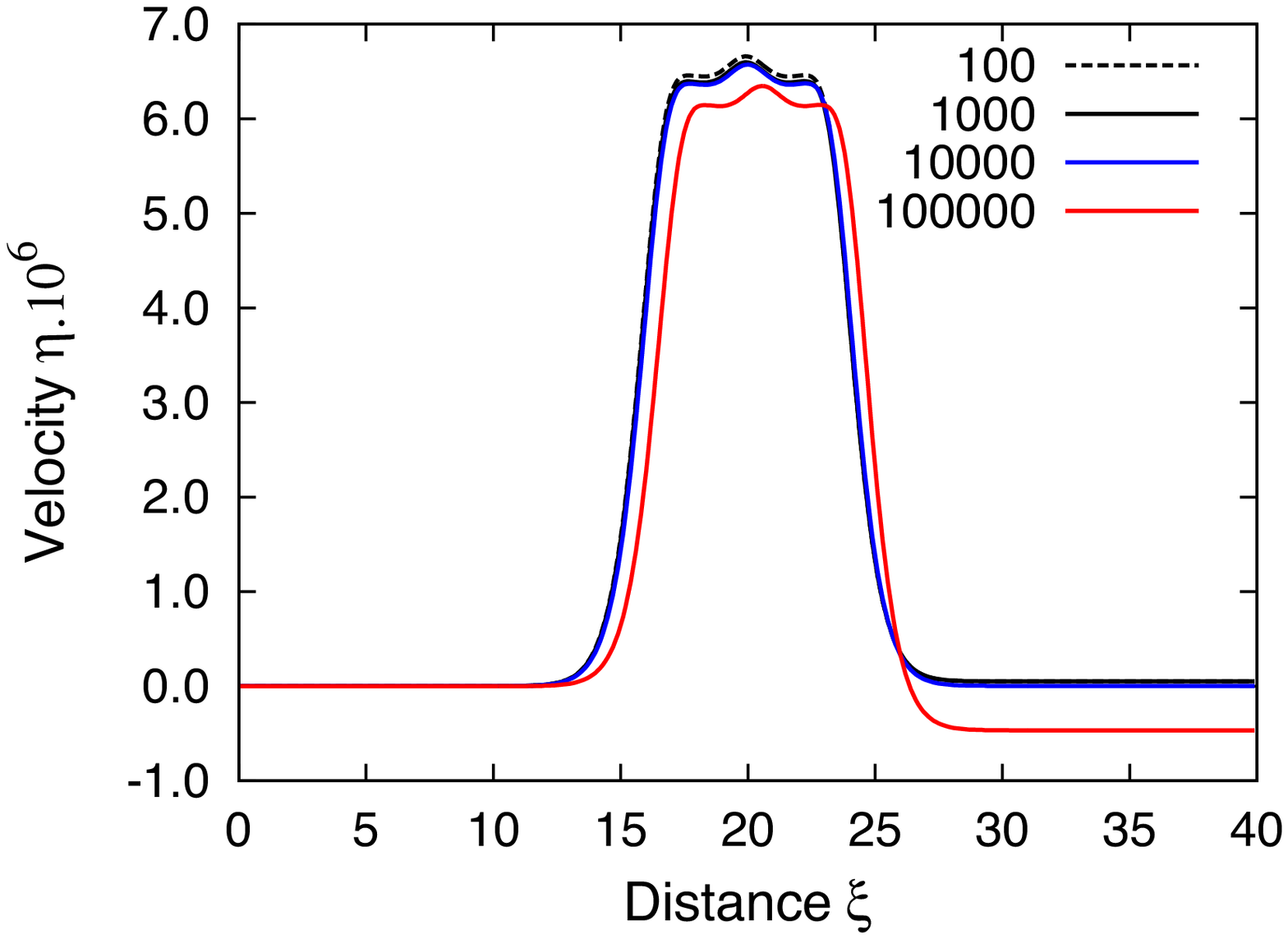}\vspace{0.12in}

\par\end{centering}

\begin{centering}
\textbf{(b)}\enskip{}\includegraphics[clip,scale=0.5]{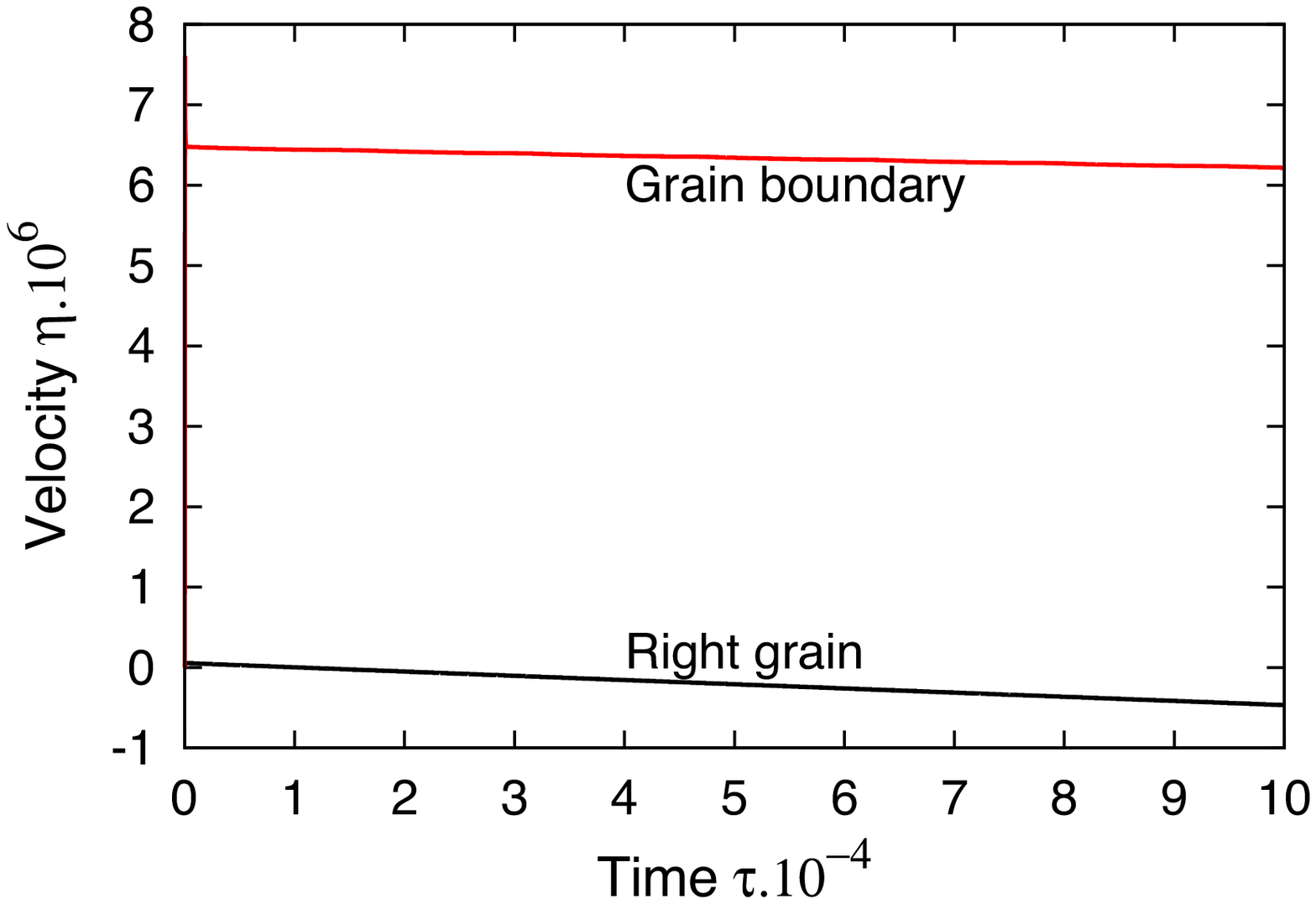}\vspace{0.12in}

\par\end{centering}

\caption{(a) Dimensionless profiles of the lattice velocity $\eta$ and (b)
the GB velocity $\eta_{GB}$ and the velocity of the right grain as
functions of time $\tau$. In (a), the time $\tau$ is indicated in
the legends. The model parameters are $\beta_{\varphi}=1$, $\beta_{s}=0.8$,
$a_{\sigma}=0$ and $a_{w}=0.5$. \label{fig:Fig-4-1}}
\end{figure}

Due to the concentration gradient, the vacancies are initially over-saturated
on the right of the GB and under-saturated on the left. To approach
equilibrium, excess vacancies must be eliminated by the GB on its
right and generated on its left. This process is accompanied by elimination
of crystal planes on the right and creation of new crystal planes
on the left. As a result, locally within the GB region, the left grain
grows while the right grain shrinks, causing GB migration to the right.
This site generation/annihilation process results in the positive
peak of the lattice velocity in the GB region (Fig.~\ref{fig:Fig-4-1}).
The small bump near the center of the peak is a non-classical effect
which originates from the deviation of the system from phase field
equilibrium {[}the term multiplying $a_{w}$ in Eq.(\ref{eq:154}){]}.
The fact that the right grain has a negative velocity indicates that
the net vacancy balance is slightly shifted towards annihilation.
It is also observed that the height of the velocity peak decreases
with time and drifts to the right together with the GB. 

This example demonstrates an interesting effect in which a GB can
be moved by a trans-gradient of vacancy concentration, a phenomenon
which could be observable experimentally. To provide an additional
proof of this effect, the calculation was repeated with the opposite
sign of the vacancy concentration gradient but the same boundary values
of the phase field. As expected, the gradient caused the GB to migrate
to the left with a nearly identical magnitude of the velocity.

\section{Summary and conclusions\label{sec:Discussion}}

The proposed theory of creep takes classical solid-state thermodynamics\cite{Larche73,Larche_Cahn_78,Larche1985,Mullins1985}
as the starting point and generalizes it in at least two ways. First,
we have lifted the ``network constraint'' and allowed lattice sites
to be created or destroyed with a rate which can be a continuous function
of coordinates and in addition can depend on crystallographic direction.
This has been achieved by introducing two different deformation gradients
co-existing in the same material, one describing local lattice distortions
due to elastic strains, compositional strain and thermal expansion,
and the other describing the total deformation including the permanent
distortion produced by the creation and annihilation of lattice sites.
The difference between the two represents the amount of creep deformation.
Accordingly, its time derivative $\mathbf{R}$ defined by Eq.(\ref{eq:24-1})
is identified with the creep deformation rate. Similar to recent work\cite{Svoboda2006,Fischer2011}
and by contrast to other creep theories, the creep deformation rate
is a tensor that encapsulates both the volume tension and compression
due to the net production or elimination of vacancies, and pure shear
deformation by concurrent site generation and annihilation without
altering the total number of sites. 

The particular formulation of the theory presented in this paper relies
on the assumption of a substitutional solid solution with a Bravais
lattice. Accordingly, for a heterogeneous material its phases are
assumed be ``coherent'' with each other, i.e., derivable from the
same reference structure by affine distortions. Furthermore, our treatment
of the deformation gradient $\tilde{\mathbf{F}}$ as a continuous
function of coordinates implies that interfaces between the phases
are coherent. In the future, this version of the theory can be generalized
to solids with interstitials and non-Bravais lattices, permitting
a more general treatment of the structures of the phases and inter-phase
interfaces. 

The tensor $\mathbf{R}$ reflects the symmetry of the material's microstructure
and the operation of particular site generation mechanisms. We gave
a few examples in which some of the components of $\mathbf{R}$ are
identically zero due to the absence of certain site generation mechanisms
or presence of geometric restrictions. In such cases, the material
can be capable of supporting static shear stresses and can reach a
(constrained) thermodynamic equilibrium in a non-hydrostatic state
of stress. When $\mathbf{R}=\mathbf{0}$, the theory reduces to the
formulation in which the solid is subject to the ``network constraint''.
If all components of $\mathbf{R}$ are nonzero, the equilibrium state
has to be hydrostatic. The ultimate equilibrium state of the material
is uniform isotropic tension or compression.

The second generalization is the addition of phase fields and their
gradients, along with gradients of concentrations of the chemical
species. Owing to this non-classical character, the kinetic equations
of the theory can automatically describe the evolution of microstructure
as part of the creep deformation process, eliminating the need to
prescribe a particular distribution of vacancy sinks and sources.
For example, the site creation and annihilation can be localized in
GBs by appropriate choice of the phase-field dependence of the kinetic
coefficient controlling the site generation rate. If the GB moves,
the vacancy sinks and sources will move together with it. 

The entropy production rate derived herein identifies several dissipation
mechanisms in the material: conduction of heat, diffusion of chemical
species, evolution of the phase fields, viscous dissipation (e.g.,
by phonons), and finally site creation and annihilation. It also identifies
the generalized forces and generalized fluxes corresponding to different
dissipation mechanisms. It particular, the creep deformation rate
$\mathbf{R}$ is identified as one such flux and the thermodynamic
force driving the creep deformation is found to be $\left(\tilde{\mathbf{\boldsymbol{\sigma}}}-\omega\mathbf{I}\right)/T$,
where $\omega$ is the non-classical grand potential density and $\tilde{\mathbf{\boldsymbol{\sigma}}}$
is the classical recoverable stress tensor. Diffusion is driven by
gradients of the non-classical diffusion potentials $M_{i}^{*}$ and
viscous dissipation by the deviation of the dynamic stress tensor
$\boldsymbol{\sigma}$ from the non-classical (Korteweg) stress $(\tilde{\mathbf{\boldsymbol{\sigma}}}+\mathbf{A}^{*})$.
The latter gives rise to interface stresses, which are thus automatically
included in this theory.

In formulating phenomenological relations between the fluxes and forces
we take into account the symmetry properties of the material.\cite{De-Groot1984}
The symmetry analysis is prepared by partitioning the entropy production
into groups of terms with the same tensor character. The existence
or absence of coupling between different groups is established by
analyzing the effect of the symmetry operations on the terms with
a particular tensor character. The case of a fully isotropic material
is analyzed in greatest detail. The splitting of the creep deformation
rate $\mathbf{R}$ into the volume and shear components emerges as
a result of this coupling analysis, with each component driven by
a different thermodynamic force. The case of axial symmetry is also
discussed as an example of less symmetric materials. In this case,
the volume and shear components of $\mathbf{R}$ are inseparable and
merge into a single tensile deformation rate $R_{n}=\mathbf{n}\cdot\mathbf{R}\cdot\mathbf{n}$,
where $\mathbf{n}$ is the unit vector parallel to the axis of symmetry.
Rigorous analysis of other symmetries relevant to particular classes
of materials would be an interesting direction for future work.

The obtained phenomenological equations can be used for formulating
a set of kinetic equations describing the evolution of the material
during creep deformation. This requires input in the form of a thermodynamic
equation of state, coordinate and time dependencies of the kinetic
coefficients and other specific properties of the material. While
this theory awaits applications to real materials, it is illustrated
in this paper by a simple one-dimensional example of a bicrystal with
a GB acting as a sink and source of vacancies. The kinetic equations
have been formulated and solved numerically for three different cases.
The calculations demonstrate how the vacancy generation or absorption
due to deviations from vacancy equilibrium or caused by applied stresses
can induce not only creep deformation of the sample but also GB migration
(moving vacancy sink/source). The calculations also reveal an interesting
effect of GB motion induced by a vacancy concentration gradient across
the boundary. This trans-gradient induced GB migration might occur
in processes such as radiation creep and deserves further study in
the future.\vspace{0.15in}

\textbf{Acknowledgement:} We are grateful to G.~B.~McFadden, J.~E.~Guyer
and J~.Ovdquist for helpful discussions in the course of this research.
This work was supported by the National Institute of Standards and
Technology, Materials Measurement Laboratory, the Materials Science
and Engineering Division.


\newpage{}\clearpage{}

\newpage{}

\newpage{}\clearpage{}

\appendix

\section{Exact momentum and energy balance relations}

In this Appendix we derive the exact momentum and energy balance relations
with respect to the lattice.

\subsection{Momentum balance}

Consider a fixed control region of a volume $V$. The rate of the
total linear momentum $\mathbf{P}$ of the region is
\begin{equation}
\dfrac{d\mathbf{P}}{dt}=\intop_{V}\dfrac{\partial}{\partial t}\left(\rho\mathbf{v}\right)dV+\intop_{\partial V}\sum_{i=1}^{n}m_{i}n_{i}\mathbf{v}_{i}\left(\mathbf{v}_{i}\cdot\mathbf{n}\right)dA,\label{eq:1}
\end{equation}
where the second integral represent the momentum dissipation through
the boundaries. Here, $\mathbf{n}$ is a unit normal pointing outside
the region and $dA$ is an increment of area of the boundary. All
other notations have been defined in the main text. Applying the divergence
theorem, 
\begin{equation}
\dfrac{d\mathbf{P}}{dt}=\intop_{V}\left[\dfrac{\partial}{\partial t}\left(\rho\mathbf{v}\right)+\nabla\cdot\mathbf{Z}\right]dV,\label{eq:3}
\end{equation}
where
\begin{equation}
\mathbf{Z}\equiv\sum_{i=1}^{n}m_{i}n_{i}\mathbf{v}_{i}\mathbf{v}_{i}=\rho\mathbf{v}_{L}\mathbf{v}_{L}+\mathbf{v}_{L}\boldsymbol{q}+\boldsymbol{q}\mathbf{v}_{L}+\mathbf{M}.\label{eq:5-3}
\end{equation}
On the other hand, the total force acting on the region is
\begin{equation}
\mathbf{F}=\intop_{V}\mathbf{b}dV+\intop_{\partial V}\mathbf{n}\cdot\boldsymbol{\sigma}dA=\intop_{V}\left(\mathbf{b}+\nabla\cdot\boldsymbol{\sigma}\right)dV.\label{eq:11}
\end{equation}
Writing down the Newton law $d\mathbf{P}/dt=\mathbf{F}$ we obtain
the momentum balance equation in the barycentric formulation, 
\begin{equation}
\dfrac{\partial}{\partial t}\left(\rho\mathbf{v}\right)+\nabla\cdot\mathbf{Z}=\mathbf{b}+\nabla\cdot\boldsymbol{\sigma}.\label{eq:6}
\end{equation}

Using the mass conservation relation
\begin{equation}
\dfrac{\partial\rho}{\partial t}=-\nabla\cdot\left(\rho\mathbf{v}\right)\label{eq:7}
\end{equation}
it can be shown that
\begin{equation}
\dfrac{\partial}{\partial t}\left(\rho\mathbf{v}\right)=\rho\dfrac{d^{L}\mathbf{v}_{L}}{dt}+\dfrac{d^{L}\boldsymbol{q}}{dt}-\nabla\cdot\left(\rho\mathbf{v}_{L}\mathbf{v}_{L}\right)-\mathbf{v}_{L}\nabla\cdot\boldsymbol{q}-\mathbf{v}_{L}\cdot\nabla\boldsymbol{q}.\label{eq:7-1}
\end{equation}
Combining the above equations, we obtain the momentum balance equation
in the lattice representation
\begin{equation}
\rho\dfrac{d^{L}\mathbf{v}_{L}}{dt}=\mathbf{b}+\nabla\cdot\left(\boldsymbol{\sigma}-\mathbf{M}\right)-\dfrac{d^{L}\boldsymbol{q}}{dt}-\boldsymbol{q}\nabla\cdot\mathbf{v}_{L}-\boldsymbol{q}\cdot\nabla\mathbf{v}_{L}.\label{eq:15-1}
\end{equation}

\subsection{Energy balance}

The internal energy density $u$ was defined in the main text through
the total energy ansatz (\ref{eq:28}). This internal energy appears
in  De Groot and Mazur\cite{De-Groot1984}, where it is denoted $u^{*}$,
but most of their discussion is focused on an approximate internal
energy defined by replacing $K$ by only the barycentric kinetic energy
$\rho|\mathbf{v}|^{2}/2$. Although this approximation greatly simplifies
all equations, our calculations will be based on the full kinetic
energy $K$.

The energy balance is governed by the energy conservation law,
\begin{equation}
\dfrac{\partial e^{\prime}}{\partial t}+\nabla\cdot\left(e^{\prime}\mathbf{v}_{L}+\mathbf{J}_{u}^{L}+\mathbf{J}_{K}^{L}\right)=\dot{w},\label{eq:29}
\end{equation}
where $\dot{w}$ is the rate of mechanical work per unit volume, $e^{\prime}\equiv e-\psi$
is the energy without the potential energy since the latter is already
included in $\dot{w}$, $\mathbf{J}_{u}^{L}$ is the internal energy
flux and $\mathbf{J}_{K}^{L}$ is the kinetic energy flux relative
to the lattice. The internal energy rate
\begin{equation}
\dfrac{d^{L}u}{dt}+u\nabla\cdot\mathbf{v}_{L}=-\nabla\cdot\left(\mathbf{J}_{u}^{L}+\mathbf{J}_{K}^{L}\right)+\dot{w}-\left(\dfrac{d^{L}K}{dt}+K\nabla\cdot\mathbf{v}_{L}\right)\label{eq:30}
\end{equation}
can be derived by computing the terms appearing in the right-hand
side. 

To compute $\dot{w}$, we consider a material region bounded by a
set of moving markers. The total work rate on this region includes
the work of volume forces and the work of stress acting on its boundary,
\begin{eqnarray}
\dot{W} & = & \intop_{V}\sum_{i=1}^{n}n_{i}\mathbf{b}_{i}\cdot\mathbf{v}_{i}dV+\intop_{\partial V}\mathbf{n}\cdot\boldsymbol{\sigma}\cdot\mathbf{v}_{L}dA\nonumber \\
 & = & \intop_{V}\left[\sum_{i=1}^{n}n_{i}\mathbf{b}_{i}\cdot\mathbf{v}_{i}+\nabla\cdot\left(\boldsymbol{\sigma\cdot\mathbf{v}_{L}}\right)\right]dV,\label{eq:20-1}
\end{eqnarray}
from which
\begin{equation}
\dot{w}=\sum_{i=1}^{n}\mathbf{b}_{i}\cdot\mathbf{J}_{i}^{L}+\mathbf{v}_{L}\cdot\left(\mathbf{b}+\nabla\cdot\boldsymbol{\sigma}\right)+\boldsymbol{\sigma}:\nabla\mathbf{v}_{L}\cdot\label{eq:22-2}
\end{equation}

The kinetic energy (\ref{eq:38}) can be split into three terms,
\begin{equation}
K=K_{L}+K_{d}+\boldsymbol{q}\cdot\mathbf{v}_{L},\label{eq:40}
\end{equation}
where 

\begin{equation}
K_{L}=\dfrac{1}{2}\rho|\mathbf{v}_{L}|^{2}\label{eq:40-1}
\end{equation}
is the macroscopic kinetic energy of the lattice motion and
\begin{equation}
K_{d}=\sum_{i=1}^{n}\dfrac{m_{i}}{2n_{i}}\mathbf{J}_{i}^{L}\cdot\mathbf{J}_{i}^{L}=\dfrac{1}{2}\textrm{Tr}(\mathbf{M})\label{eq:18}
\end{equation}
is the kinetic energy of diffusion. Calculations show that 
\begin{equation}
\dfrac{d^{L}K_{L}}{dt}+K_{L}\nabla\cdot\mathbf{v}_{L}=\rho\dfrac{d^{L}\mathbf{v}_{L}}{dt}\cdot\mathbf{v}_{L}-\dfrac{1}{2}|\mathbf{v}_{L}|^{2}\nabla\cdot\boldsymbol{q},\label{eq:43}
\end{equation}
\begin{equation}
\dfrac{d^{L}K_{d}}{dt}+K_{d}\nabla\cdot\mathbf{v}_{L}=-\sum_{i=1}^{n}\dfrac{m_{i}}{2n_{i}^{2}}\left(\mathbf{J}_{i}^{L}\cdot\mathbf{J}_{i}^{L}\right)\nabla\cdot\mathbf{J}_{i}^{L}+\sum_{i=1}^{n}m_{i}\mathbf{J}_{i}^{L}\cdot\dfrac{d^{L}\mathbf{w}_{i}}{dt},\label{eq:35-3}
\end{equation}
\begin{equation}
\dfrac{d^{L}}{dt}\left(\boldsymbol{q}\cdot\mathbf{v}_{L}\right)+\left(\boldsymbol{q}\cdot\mathbf{v}_{L}\right)\left(\nabla\cdot\mathbf{v}_{L}\right)=\boldsymbol{q}\cdot\dfrac{d^{L}\mathbf{v}_{L}}{dt}+\mathbf{v}_{L}\cdot\dfrac{d^{L}\boldsymbol{q}}{dt}+\left(\boldsymbol{q}\cdot\mathbf{v}_{L}\right)\left(\nabla\cdot\mathbf{v}_{L}\right),\label{eq:42}
\end{equation}
where we denoted $\mathbf{w}_{i}\equiv\mathbf{v}_{i}-\mathbf{v}_{L}$.
Summing up Eqs.(\ref{eq:43}), (\ref{eq:35-3}) and (\ref{eq:42}),
we obtain kinetic energy rate
\begin{eqnarray}
\dfrac{d^{L}K}{dt}+K\nabla\cdot\mathbf{v}_{L} & = & \left(\rho\mathbf{v}_{L}+\boldsymbol{q}\right)\cdot\dfrac{d^{L}\mathbf{v}_{L}}{dt}+\mathbf{v}_{L}\cdot\dfrac{d^{L}\boldsymbol{q}}{dt}+\left(\boldsymbol{q}\cdot\mathbf{v}_{L}\right)\left(\nabla\cdot\mathbf{v}_{L}\right)-\dfrac{1}{2}|\mathbf{v}_{L}|^{2}\nabla\cdot\boldsymbol{q}\nonumber \\
 &  & -\sum_{i=1}^{n}\dfrac{m_{i}}{2n_{i}^{2}}\left(\mathbf{J}_{i}^{L}\cdot\mathbf{J}_{i}^{L}\right)\nabla\cdot\mathbf{J}_{i}^{L}+\sum_{i=1}^{n}\dfrac{m_{i}}{n_{i}}\mathbf{J}_{i}^{L}\cdot\dfrac{d^{L}\mathbf{w}_{i}}{dt}.\label{eq:44}
\end{eqnarray}
For the lattice flux of kinetic energy we have
\begin{eqnarray}
\mathbf{J}_{K}^{L} & = & \sum_{i=1}^{n}\dfrac{1}{2}m_{i}n_{i}\left(\mathbf{v}_{i}\cdot\mathbf{v}_{i}\right)\left(\mathbf{v}_{i}-\mathbf{v}_{L}\right)\nonumber \\
 & = & \sum_{i=1}^{n}\dfrac{m_{i}}{2n_{i}^{2}}\left(\mathbf{J}_{i}^{L}\cdot\mathbf{J}_{i}^{L}\right)\mathbf{J}_{i}^{L}+\dfrac{1}{2}\boldsymbol{q}|\mathbf{v}_{L}|^{2}+\mathbf{M}\cdot\mathbf{v}_{L}.\label{eq:50-1-1}
\end{eqnarray}

Inserting Eqs.(\ref{eq:22-2}), (\ref{eq:44}) and (\ref{eq:50-1-1})
in the right-hand side of Eq.(\ref{eq:30}), after lengthy calculations
we finally obtain
\begin{eqnarray}
\dfrac{d^{L}u}{dt}+u\nabla\cdot\mathbf{v}_{L} & = & -\nabla\cdot\mathbf{J}_{u}^{L}+\sum_{i=1}^{n}\mathbf{b}_{i}\cdot\mathbf{J}_{i}^{L}+\left(\boldsymbol{\sigma}-\mathbf{M}\right):\nabla\mathbf{v}_{L}\nonumber \\
 &  & -\sum_{i=1}^{n}\left\{ \nabla\left[\dfrac{m_{i}}{2n_{i}^{2}}\left(\mathbf{J}_{i}^{L}\cdot\mathbf{J}_{i}^{L}\right)\right]+m_{i}\dfrac{d^{L}\mathbf{v}_{i}}{dt}\right\} \cdot\mathbf{J}_{i}^{L}.\label{eq:50-2}
\end{eqnarray}

\section{Justification of the approximate form of the entropy production}

In the main text, we derived the exact expression for the entropy
production rate (\ref{eq:84-1}). Deriving the linear constitutive
relations we assumed that the fluxes and forces were both small. Under
this linear approximation, all terms quadratic in fluxes and/or forces
must be neglected. Tensor $\mathbf{M}$ defined by Eq.(\ref{eq:13-2})
is quadratic in the diffusion fluxes $\mathbf{J}_{i}^{L}$ and can
be neglected. Further, the term $m_{i}(\mathbf{J}_{i}\cdot\mathbf{J}_{i})/2n_{i}^{2}$
in the driving force of diffusion is also quadratic in diffusion fluxes
and can also be neglected. 

Furthermore, for slow processes such as creep the inertia terms $m_{i}d^{L}\mathbf{v}_{i}/dt$
can be neglected after a short transient. To demonstrate this, consider
an isotropic materials not subject to external fields. Assuming a
uniform temperature field, diffusion is decoupled from all other processes
and is described by the equations 
\begin{equation}
\mathbf{J}_{i}^{L}=-\dfrac{1}{T}\sum_{j=1}^{n}L_{ij}\left(\nabla M_{j}^{*}+m_{j}\dfrac{d^{L}\mathbf{v}_{j}}{dt}\right),\enskip\enskip i=1,...,n\label{eq:56}
\end{equation}
where the $n\times n$ matrix of kinetic coefficients $\mathbf{L}$
is symmetric and positive definite. 

Rewrite (\ref{eq:56}) 
\begin{equation}
\mathbf{J}_{i}^{L}=-\dfrac{1}{T}\sum_{j=1}^{n}L_{ij}\left(\nabla M_{j}^{*}+m_{j}\dfrac{d^{L}\mathbf{w}_{j}}{dt}+m_{j}\dfrac{d^{L}\mathbf{v}_{L}}{dt}\right),\enskip\enskip i=1,...,n\label{eq:57}
\end{equation}
and consider the effect of each inertia term separately. To understand
the role of the first inertia term, suppose all other driving forces
are zero. Neglecting also the cross-effects among the diffusion fluxes,
the diffusion equations reduce to
\begin{equation}
\mathbf{w}_{i}=-\dfrac{m_{i}L_{ii}}{Tn_{i}}\dfrac{d^{L}\mathbf{w}_{i}}{dt}\equiv-\tau_{i}\dfrac{d^{L}\mathbf{w}_{i}}{dt},\enskip\enskip i=1,...,n.\label{eq:58}
\end{equation}
Assuming that
\begin{equation}
\tau_{i}=\dfrac{m_{i}L_{ii}}{Tn_{i}}\label{eq:59}
\end{equation}
is a slow-varying function of time, Eqs.(\ref{eq:58}) have approximately
exponential solutions $\mathbf{w}_{i}\propto\exp(-t/\tau_{i})$ showing
that any initial acceleration of the particles relative to the lattice
damps after a characteristic time $\tau_{i}$. Thus, for processes
occurring on time scales much longer than $\tau_{i}$, the inertia
terms $m_{j}d^{L}\mathbf{w}_{j}/dt$ can be neglected.

To evaluate typical values of $\tau$, take one of the species, say
1, and express the kinetic coefficient $L_{11}$ through the diffusion
coefficient $D_{1}$ via $L_{11}=n_{1}D_{1}/k_{B}$, which gives $\tau=m_{1}D_{1}/(k_{B}T)$.
Taking the molecular weight of 100 a.m.u., the upper bound of the
diffusion coefficients in solids $D_{1}=10^{-9}$ m$^{2}$/s and the
temperature of 1000 K we obtain $\tau\approx10^{-14}$ s. At lower
temperatures $\tau$ is even smaller. Thus the time scale of damping
of the inertia terms is much smaller than the typical time scale of
creep tests (many hours). 

The inertia terms $m_{i}d^{L}\mathbf{v}_{L}/dt$ originate from the
accelerated lattice motion due to applied mechanical stress as well
as the site generation and other relatively slow processes. Before
the material reaches mechanical equilibrium, the lattice velocities
can be very high, possibly comparable with the speed of sound, and
the inertia force $m_{i}d^{L}\mathbf{v}_{L}/dt$ can be significant.
But the subsequent creep deformation is a slow process in which the
material maintains mechanical equilibrium and $d^{L}\mathbf{v}_{L}/dt$
reflects only the slow changes in the creep deformation rate. As a
crude estimate, the magnitude of the lattice acceleration is related
to variations in the creep deformation $\dot{\epsilon}$ by
\begin{equation}
\dfrac{d^{L}\ln|\mathbf{v}_{L}|}{dt}\approx\dfrac{d\ln\dot{\epsilon}}{dt}.\label{eq:60}
\end{equation}

During the steady-state creep, $\dot{\epsilon}$ remains nearly constant
and depends only on the applied stress and temperature, so that the
inertia terms $m_{i}d^{L}\mathbf{v}_{L}/dt$ can be neglected. During
the primary and tertiary stages, the right-hand side of Eq.(\ref{eq:60})
still remains small. For example, typical steady-state creep rates
in metallic alloys are $\dot{\epsilon}\approx10^{-6}$ to $10^{-3}$
s$^{-1}$. During the primary and tertiary stages, the rate changes
by at most an order of magnitude over hundreds of hours. Thus, as
an upper bound $\ddot{\epsilon}\approx10^{-7}$ s$^{-2}$ and thus
$d^{L}\ln|\mathbf{v}_{L}|/dt\approx0.1$ s$^{-1}$. 

To show that the inertia effects are negligible, we combine the particle
conservation law with the diffusion equation for species $i$, 
\begin{equation}
\mathbf{J}_{i}^{L}=-\dfrac{1}{T}L_{ii}\left(\nabla M_{j}^{*}+m_{i}\dfrac{d^{L}\mathbf{v}_{L}}{dt}\right),\label{eq:62}
\end{equation}
in which we again neglected the cross effects among the fluxes. Treating
the kinetic coefficient as a constant,
\begin{equation}
\dfrac{\partial n_{i}}{\partial t}=-\dfrac{1}{T}L_{ii}\nabla^{2}M_{j}^{*}-\nabla\cdot\left(n_{i}\mathbf{v}_{L}-\dfrac{1}{T}L_{ii}m_{i}\dfrac{d^{L}\mathbf{v}_{L}}{dt}\right).\label{eq:63}
\end{equation}
The ratio of the second term to the first inside the divergence is
on the order of
\begin{equation}
\dfrac{L_{ii}m_{i}}{Tn_{i}}\dfrac{d^{L}\ln|\mathbf{v}_{L}|}{dt}=\tau_{i}\dfrac{d^{L}\ln|\mathbf{v}_{L}|}{dt},\label{eq:64}
\end{equation}
where $\tau_{i}$ is the characteristic time (\ref{eq:59}). The latter
was estimated to be $\sim10^{-14}$ s. Thus, the inertia term in Eq.(\ref{eq:63})
is more than ten orders of magnitude smaller than the normal term
$n_{i}\mathbf{v}_{L}$, reducing the diffusion equation to the usual
form
\begin{equation}
\dfrac{\partial n_{i}}{\partial t}=-\dfrac{1}{T}L_{ii}\nabla^{2}M_{j}^{*}-\nabla\cdot\left(n_{i}\mathbf{v}_{L}\right).\label{eq:65}
\end{equation}

These estimates justify the approximate form (\ref{eq:84}) of the
entropy production for creep applications.
\end{document}